\documentclass{article}

\usepackage{arxiv}

\usepackage[utf8]{inputenc} % allow utf-8 input
\usepackage[T1]{fontenc}    % use 8-bit T1 fonts
\usepackage{hyperref}       % hyperlinks
\usepackage{amsmath}
\usepackage{amsfonts}
\usepackage{amssymb}
\usepackage{eufrak}
\usepackage{eucal}
\usepackage{braket}
\usepackage{apacite}
\usepackage{natbib}
\usepackage{url}            % simple URL typesetting
\usepackage{booktabs}       % professional-quality tables
\usepackage{amsfonts}       % blackboard math symbols
\usepackage{nicefrac}       % compact symbols for 1/2, etc.
\usepackage{microtype}      % microtypography
\usepackage{cleveref}       % smart cross-referencing
\usepackage{lipsum}         % Can be removed after putting your text content
\usepackage{graphicx}
\usepackage{natbib}
\usepackage{multirow}
\usepackage{doi}
\usepackage{diagbox}
\usepackage{appendix}
\usepackage{epigraph}
\usepackage{algorithm}
\usepackage{setspace}
\usepackage{upgreek}
\usepackage{algpseudocode}% http://ctan.org/pkg/algorithmicx
\usepackage{xcolor}
\usepackage{bm}

\algnewcommand{\LineComment}[1]{\State \(\triangleright\) #1}
\usepackage{etoolbox}
\usepackage{tikz}
\usetikzlibrary{arrows,positioning} 

\tikzset{every picture/.style={line width=0.75pt}} %set default line width to 0.75pt        

\makeatletter
\patchcmd{\epigraph}{\@epitext{#1}}{\itshape\@epitext{#1}}{}{}
\makeatother

\title{A communication \emph{Satellite} Services based Dencentralized Network Protocol}

\newif\ifuniqueAffiliation
\uniqueAffiliationtrue

\ifuniqueAffiliation % Standard variant of author block

\author{
	Xiao Yan\\
	\texttt{shiotoli@gmail.com}
	\And
	Bernie Gao \\
	\texttt{bgao150@aucklanduni.ac.nz}
}
\else
\author{Xiao Yan (\texttt{shiotoli@gmail.com}) \AND Bernie Gao (\texttt{bbb@aaa.com})}
\fi

\renewcommand{\headeright}{\textit{Ver.} 0.2.0}
\renewcommand{\undertitle}{Technical White Paper}
\renewcommand{\shorttitle}{Space Network Protocol}
\DeclareMathOperator*{\argmax}{arg\,max}

\hypersetup{
pdftitle={A communication Satellite Services based Dencentralized Network Protocol},
pdfsubject={q-bio.NC, q-bio.QM},
pdfauthor={Xiao Yan},
pdfkeywords={Communication Satellite, Blockchain, Decentralized Network},
}
\newtheorem{theorem}{Theorem}[section]

\newtheorem{hypothesis}[theorem]{Hypothesis}

\usepackage{todonotes}
\algnewcommand\algorithmicforeach{\textbf{for each}}
\algdef{S}[FOR]{ForEach}[1]{\algorithmicforeach\ #1\ \algorithmicdo}

\begin{document}
\maketitle
\begin{abstract}
% 在本文中我们提出了一种基于通信卫星服务的去中心化网络协议——(CSP)。通过该协议，该协议定义了如何将通信卫星的网络服务的情况发放至整个去区块链网络中，使得针对于所有的通信服务更加的公平和透明。
% 我们的核心目标是将所有卫星网络所提供的服务对于无论卫星或者终端硬件都存在不可靠风险的情况下，都可以根据CSP协议达成一致性共识。为了实现这个目标，我们提出了PoD(Proof of Onbroad)的概念。
% 通过零知识证明和多方加密计算的方式，针对终端节点或者网络节点欺诈的情况下，依然可以评估出每一颗卫星的服务提供参数。通过该项技术可以给分布式的网络服务模型提供技术支持。
In this paper, we present a decentralized network protocol, \emph{Space Network Protocol}, based on Communication Satellite Services. The protocol outlines a method for distributing information about the status of satellite communication services across the entire blockchain network, facilitating fairness and transparency in all communication services.
Our primary objective is to standardize the services delivered by all satellite networks under the communication satellite protocol. This standard remains intact regardless of potential unreliability associated with the satellites or the terminal hardware. We proposed PoD (Proof of Distribution) to verify if the communication satellites are online and PoF (Proof of Flow) to authenticate the actual data flow provided by the communication satellites. In addition, we also proposed PoM (Proof of Mesh) to verify if the communication satellites have successfully meshed together.
Utilizing zero-knowledge proof and multi-party cryptographic computations, we can evaluate the service provisioning parameters of each satellite, even in the presence of potential terminal or network node fraud. This method offers technical support for the modeling of distributed network services.

\end{abstract}

% keywords can be removed

\section{Introduction}

% \epigraph{My brain is only a receiver, in the Universe there is a core from which we obtain knowledge, strength and inspiration. I have not penetrated into the secrets of this core, but I know that it exists.}{--- \textup{Nikola Tesla}}
As we delve into the 21st century, technological advancements have propelled the evolution of communication networks. Satellite communication, in particular, has significantly expanded in both scope and capabilities. Communication satellites have transformed from being used primarily for TV broadcasting or meteorological data collection to becoming integral for global communication networks, enabling connectivity in remote and underserved areas. They are now indispensable in crucial sectors such as defense, disaster management, navigation, and even internet provision. 

Indeed, the move of businesses such as Starlink towards commercial satellite communication coverage represents a significant development in global connectivity. Projects like these aim to bridge the digital gap, providing even the remotest regions with access to the internet. Consequently, this progress has essential implications for global inclusivity and equality of information access. By providing satellite network services, these companies are contributing significantly to worldwide efforts for comprehensive internet availability, thereby playing a crucial role in promoting social and economic development across the globe. 

Absolutely, an objective evaluation of communication satellite service quality has indeed become one of the crucial goals in technological advancement. As the industry expands and communication satellite services become more prevalent, maintaining high-quality services is paramount. This not only involves aspects such as signal coverage and transmission efficiencies but also includes reliability, data security, and user privacy. The continuous improvement of these service quality parameters underlines the need for innovative solutions and advances in technology, ultimately benefiting the industry and its users.

At the same time, the development and proliferation of blockchain technology have brought about a paradigm shift in the way we manage and store information. Initially associated solely with digital currencies such as Bitcoin\citeyearpar{bitcoin}, blockchain technology has grown beyond financial transactions. It has become a pivotal tool in multiple domains, offering secure, decentralized, and transparent methods to manage and verify transactions and exchanges of information.

Blockchain technology has remarkably transformed the landscape of digital transactions by providing a decentralized, transparent, and secure system for recording and verifying transactions. Crucial to this innovative technology is the concept of 'consensus.'

Consensus, in the context of blockchain, refers to the agreement achieved among all participants or nodes in the network regarding the validity of transactions. This agreement, determined through certain algorithms, is critical as it ensures the synchronous maintenance of the distributed ledger without requiring a central authority.

Different blockchain platforms employ varying consensus mechanisms, each presenting unique characteristics and advantages. For instance:
\begin{itemize}
	\item Proof of Work (PoW): The Bitcoin\citeyearpar{bitcoin} network uses PoW as its consensus mechanism. In PoW, miners expend computational resources to solve complex mathematical problems, which validates transactions and adds new blocks to the chain. This mechanism provides a high-security assurance at the cost of significant computational resources.
	\item Proof of Stake (PoS): The Ethereum\citeyearpar{beth,weth} network, which originally used PoW, is transitioning to a PoS consensus mechanism referred to as 'Ethereum 2.0.' In PoS, validators are chosen to create new blocks based on the number of ethers a node owns and is prepared to 'stake' to vouch for the transaction's validity. PoS aims to achieve the same security as PoW but with reduced computational and energy requirements.
	\item Byzantine Fault Tolerance (BFT): Blockchain platforms such as Cosmos\citeyearpar{cosmos} and Polkadot\citeyearpar{wood_polkadot_2016} utilize variations of the BFT consensus mechanism. Cosmos employs a form known as Tendermint BFT, while Polkadot uses GHOST-based Recursive Ancestor Deriving Prefix Agreement (GRANDPA)\citeyearpar{stewart_grandpa_2019}. These mechanisms, designed to tackle the Byzantine Generals Problem, provide strong fault tolerance even in the presence of rogue or faulty nodes, enhancing the robustness and security of the networks.
\end{itemize}

Each of these consensus mechanisms plays a critical role in maintaining the trust, security, and decentralization intrinsic to blockchain technology. As such, they serve as the technological backbone enabling the versatile applications of blockchain in wide-ranging sectors.

\subsection {Motivation}
In the modern era of technological advancements, communication satellite network services can greatly benefit from the incorporation of blockchain technology; particularly, the concept of consensus can be applied to ascertain service parameters and ensure network integrity.

Our core objective is to achieve consensus on three critical aspects of communication satellite services, each verified through unique Proof mechanisms.
\begin{itemize}
	\item \textbf{Network Meshing}: This section verifies whether the current satellite has successfully integrated into the network. For this, we propose "Proof of Mesh" (PoM) in Sec.~\ref{sec:pom} which certifies the establishment of network connections. Furthermore, PoM not only verifies whether a satellite has successfully joined the network, but also implements it through a key exchange protocol.
	\item \textbf{Satellite Online Status}: To confirm whether the satellite is successfully online and operational, we introduce "Proof of Distribution" (PoD) in Sec.\ref{sec:pod}. This proof acts as a verifiable "heartbeat" of the satellite, affirming its operational status. PoD primarily operates on two assumptions—time continuity assumption and space continuity assumption, detailed in Sec.\ref{sec:space_time_con_hypo}.
	\item \textbf{Service Flow Rate}: To validate the actual traffic flow services provided by the satellite, we propose "Proof of Flow" (PoF) in Sec.~\ref{sec:pof}. This mechanism quantitatively evaluates the service provision parameters, certifying the volume of data a satellite is capable of handling at a given time.
\end{itemize}
By leveraging these distinct Proof mechanisms in conjunction with blockchain technology, we aim to enhance the reliability, fairness, and transparency of communication satellite services. These novel solutions promise to define a new standard in the field, driving the future trajectory for satellite network services.

In Sec.~\ref{sec:consensus}, we introduce the consensus framework of the Space Network, which ensures the proper functioning of the entire blockchain network through a system of rewards and slashing. To achieve this goal, we define various consensus roles within the network, including challengers, validators, and leaders.

\subsection{Basic Definition}
\label{sec:basic_def}
In this section, we present definitions for various mathematical notations to enhance the reader’s understanding. Participants in the network are outlined as a set of transmitter, denoted by $\Set{\mathcal{T}{\star}}$, and a set of receivers, denoted by $\Set{\mathcal{R}{\star}}$. The act of a transmitter providing a service to a receiver is represented as $\mathcal{T}\mapsto\mathcal{R}$.

State parameters such as bandwidth, latency, and online status are denoted by $\alpha^\star$, with the unified set of these parameters represented as $A = {\alpha^\star}$. These parameters symbolize instantaneous values at a specific time point, corresponding to the parameters' values within the timeframe $[T,T+\mathcal{O}(T)]$. For instance, $\alpha^\text{bandwidth}(T)$ designates the instantaneous bandwidth at the time T.
In a similar vein, cumulative data and their data sets are denoted by $\beta$ and $B$, respectively. This type of data generally signifies a function within a specific time interval, expressed here as $T+\Delta T$. For example, $\beta^F(T+\Delta T)$ represents the traffic delivered within the interval $[T,T+\Delta T]$.
Location information of $\mathcal{T}$ and $\mathcal{R}$ is typically represented by $\mathbb{P}$.

\section{Wireless Satellite Communications}
This chapter primarily addresses the wireless signal strength model, focusing particularly on the signal transmission model applicable to space satellites and terrestrial receivers. Central to the discussion are two key assumptions pivotal to consensus algorithms: spatial continuity and temporal continuity. Instead of delving deep into the intricacies of the related formulas, this chapter accords higher importance to the qualitative conclusions derived from these equations.

\subsection{Free-Space Path Loss}
Let us consider the propagation of a radio wave from a source, denoted as $\mathcal{T}$, to a receiver, denoted as $\mathcal{R}$. The model for wireless signal propagation is generally characterized by the following equation \citeyearpar{satelite_communications_book}

\begin{equation}
	\label{equation:free_space_path_loss}
p_{\mathcal{R}}=p_\mathcal{T}g_\mathcal{T}g_\mathcal{R}\frac{1}{(4\pi)^2}\frac{\lambda^2}{\lVert\mathcal{T}\mapsto\mathcal{R}\rVert}
\end{equation}

$p$ stands for power, $g$ represents gain, and $\lambda$ signifies wavelength. We decompose Eqn.~\ref{equation:free_space_path_loss} into three parts:

\begin{equation}
	\label{equation:free_space_path_loss_split}
	\begin{matrix}
		p_{\mathcal{R}}=& \underbrace{p_\mathcal{T}g_\mathcal{T}g_\mathcal{R}}&\times & \underbrace{\frac{1}{(4\pi)^2} \frac{\lambda^2}{\lVert\mathcal{T}\mapsto\mathcal{R}\rVert^2}}\\
		&\text{Intrinsic Var.}& & l_p
	\end{matrix}
\end{equation}

The intrinsic variables part refers to the inherent properties of either the transmitting or receiving devices. These parameters do not change over space and time for a stable transmission-reception system. $l_p$ is defined as the path loss, which amplifies with an increase in transmission distance. Furthermore, we observe that a longer wavelength implies a stronger signal penetration ability. We represent Eqn.~\ref{equation:free_space_path_loss_split} in the form of decibels (dB):
\begin{equation}
	\label{equation:dbm}
	\mathcal{L}_p=-20\text{log}\left(\frac{4\pi \lVert\mathcal{T}\mapsto\mathcal{R}\rVert}{\lambda}\right)
\end{equation}

$l_p$ and $\mathcal{L}_p$ are typically termed 'Free-Space Path Loss'. In other words, within an unobstructed space, signals tend to attenuate with increasing distance.

However, in reality, signal propagation does not occur under ideal conditions resembling free space due to the abundance of obstacles within the environment that impede signal transmission. Section ~\ref{sec:signal_loss} provides an in-depth discussion on other causes of signal attenuation.
\subsection{Path Transmission Impairments}
\label{sec:signal_loss}

This section discusses the potential causes of data attenuation due to noise in the link between $\mathcal{R}$ and $\mathcal{T}$. For example, communication between satellites and ground facilities may experience path loss due to atmospheric impact, while communication between ground facilities can be attenuated due to obstructions like buildings or trees. Other potential barriers may include artifact electromagnetic interference; however, we only consider objectively induced attenuation here, as such interference can be classified as intrinsic signal blocking.

The carrier signal frequency used has different sensitivities to obstructions. For example, signals below 3GHz may scatter in the ionosphere of the atmosphere. On the other hand, signals above 3GHz are primarily obstructed by barriers in the troposphere (such as rain and clouds). As space satellites generally use signal frequencies above 3GHz (C, Ku, Ka, V), the main propagation obstacles stem from the troposphere. We only discuss potential causes of attenuation here.

Noise types can be divided into two main categories:

\begin{enumerate}
	\item  \textbf{System Noise}:	The noise generated by the amplifiers in the receiving system is a primary consideration. Furthermore, devices such as oscillators, up-converters, down-converters, switches, combiners, and multiplexers all generate noise.

	As these types of noise belong to internal noise, we do not discuss them in detail here.

	\item \textbf{Path Noise}: This noise is due to the properties of the medium in which the signal travels. For satellite communication, especially high-frequency signal communication (around 10GHz), one of the primary causes of loss is attenuation due to precipitation, clouds, gas absorption, tropospheric scintillation, etc., in the troposphere and signal depolarization\citeyearpar{two_paper_satcom2}.

	Introducing the noise attenuation term $l_o$ into Eqn.~\ref{equation:free_space_path_loss}, we break $l_p$ down into $l_{fs}$ and $l_o$, where $l_p=l_{fs}l_o$, i.e., the free space term and the obstruction term.

	Translated into the form of decibels (dB), we get the following equation:

	\begin{equation}
		\mathcal{L}_p = \mathcal{L}_\text{fs} + \mathcal{L}_o
	\end{equation}
	\item \textbf{Local Environment Attenuation}: The attenuation is primarily due to the local positioning and its environmental conditions, causing multipath and shadow fading. This is mainly caused by blockages and multiple paths reaching out of phase due to refraction, scattering, and reflection. This topic is particularly addressed in Sec.~\ref{sec:local}.
\end{enumerate}

Regarding path noise, the radio wave impairment term $\mathcal{L}_o$ may arise due to various reasons:
\begin{itemize}
	\item \textbf{Atmospheric Reasons}: This includes
	  \begin{itemize}
		\item Emissions from atmospheric gases such as oxygen and water vapor.
		\item Emissions from water bodies like rain and clouds.
		\item Atmospheric noise brought about by lightning discharges.
	  \end{itemize}
	\item \textbf{Terrestrial Obstructions}: This often involves re-radiation from obstacles within the ground or within the beam of the antenna.
	\item \textbf{Extraterrestrial Obstructions}: These could include
	  \begin{itemize}
		\item Cosmic background radiation.
		\item Radiation from the sun and the moon.
		\item Radiation originated from celestial radio sources, such as radio stars.
	  \end{itemize}
	\item \textbf{Artifact Obstructions}: These include
	  \begin{itemize}
		\item Unintentional radiation from electromechanical, electrical, and electronic equipment.
		\item Power transmission lines.
		\item Ignition of internal combustion engines.
		\item Radiation from other communication systems.
	  \end{itemize}
  \end{itemize}
  
  The primary factors affecting the quality of communication between space and Earth are weather conditions such as rainfall and fog. Upon considering each potential cause for attenuation and the distribution of each type of attenuation:

\begin{equation}
	\begin{aligned}
		\label{path_loss_obstacles}
	\mathcal{L}_o &= \int_{L} \text{P}^{total}_o \text{d}l\\
				&=\int_{L}\sum_i\text{P}^i_o \text{d}l\\
				&=\int_L \left[\sum_i \int_\Omega A^i_a(l,\Omega)\eta(\Omega) \text{d}\Omega\right]\text{d}l\\
            &=\sum_i \int_L \left[ \int_\Omega A^i_a(l,\Omega)\eta(\Omega) \text{d}\Omega\right]\text{d}l\\
	\end{aligned}
\end{equation}
Here, $\text{P}$ represents the attenuation density, expressed in dB/km. $\Omega$ represents the range of values of the independent variable for the attenuation factors. $A_a$ indicates the total attenuation amplitude, and $\eta$ represents the distribution at the current $\Omega$. For instance, a classic rain attenuation model can be expressed as\citeyearpar{satelite_communications_book}:

\begin{equation}
	A = C\int A_r(r,\lambda)e^{-\gamma r}\text{d}r
	\end{equation}

In the above, $C$ is a constant, $A_r$ is the attenuation per unit cross-section, and $\gamma$ is an empirical parameter.

Path loss is generally modeled at the statistical level. In terms of atmospheric impediments, they are characterized by their relatively large scale and infrequent changes over time. The influence caused by the path doesn't only result in attenuation but also leads to high-frequency noise\citeyearpar{two_paper_satcom1}. An ideal noise model assumption is drawn from the Additive White Gaussian Noise (AWGN) channel model, which presupposes that high-frequency noise is generated in the form of white noise. i.e. $ \nu \sim \mathcal{N}(0, \sigma^2)$.

\subsubsection{Local Conditions}
\label{sec:local}
Our previous section examined the potential causes for signal degradation between satellite and terrestrial equipment communication channels, and conducted a qualitative analysis of fade forms (Eqn.~\ref{path_loss_obstacles}). This section will mainly discuss potential local signal attenuation and noise.
Since the terrestrial environment, unlike the idealized free space where it can be considered devoid of matter, the potential impacts of electromagnetic waves in a terrestrial environment must be taken into account. At a macro view, electromagnetic waves exhibit the following scenarios:
\begin{itemize}
\item \textbf{Refraction}: Refraction is a phenomenon that changes the path of electromagnetic waves as they propagate between different mediums. This occurs, for instance, when wireless signals propagate from the atmosphere into a building, or from one temperature environment to another. Refraction can result in signal attenuation and multipath propagation.
\item \textbf{Reflection}: Reflection occurs when electromagnetic waves encounter obstacles larger than their wavelength and bounce back from the interface. In wireless communication, structures such as buildings, mountains, and terrain can cause signal reflection. Reflection can cause multipath propagation, and because part of the signal's energy is absorbed by obstacles, it also results in signal attenuation.
\item \textbf{Scattering}: Scattering happens when electromagnetic waves encounter obstacles smaller than their wavelength (like dust, raindrops, etc.), causing the waves to scatter. The intensity of scattering generally depends on the frequency of the wireless signal. Scattering disturbs the initially coherent electromagnetic waves, causing attenuation and increasing multipath effects.
\item \textbf{Absorption}: Absorption occurs when electromagnetic waves are absorbed by medium materials (like buildings, trees, atmosphere, raindrops, etc.) during propagation, leading to energy loss. A significant reason for the weakening of signal energy during wireless signal propagation is absorption. Absorption primarily results in attenuation, rather than multipath effects for the signal.
\item \textbf{Diffraction}: Diffraction arises when electromagnetic waves encounter obstacles (whose size is near or larger than the wavelength), causing the waves to bend at the edges of the obstacles and enter the shadowed areas of the obstacles. This is because electromagnetic waves have the ability to "bypass" obstacles during propagation. In areas filled with terrain obstacles like buildings and mountains, the diffraction effect significantly impacts signal propagation. Diffraction can lead to multipath propagation effects and potentially cause signal attenuation.
\end{itemize}
Because of these effects, two additional components are added to the original path propagation formula Eqn.~\ref{equation:free_space_path_loss}:
\begin{equation}
\begin{matrix}
\label{equation:free_space_path_loss_shadow_multi}
p_{\mathcal{R}}=p_\mathcal{T}g_\mathcal{T}g_\mathcal{R}\frac{1}{(4\pi)^2}\frac{\lambda^2}{\lVert\mathcal{T}\mapsto\mathcal{R}\rVert}&\times&\underbrace{\alpha_{sa} e^{\alpha_{se}}}&\times&\underbrace{\alpha_{m}^2}\\
&&\text{shadow}& & \text{multipath}\\
\end{matrix}
\end{equation}
The shadow fade is a mid-frequency signal following a log-normal distribution, primarily caused by shadows from buildings or trees. Its fluctuations range within several wavelengths to tens of wavelengths,
\begin{equation}
\alpha_{se}(x;\mu,\sigma^2) \sim \frac{1}{x\sigma\sqrt{2\pi}}e^{-\frac{[\text{ln}(x)-\mu]^2}{2\sigma^2}}
\end{equation}
The multipath fade is a mid-frequency log-normal distributed signal, often fluctuating within a wavelength,
\begin{equation}
\alpha_{m}(x;\mu,\sigma^2) \sim \frac{x}{\sigma^2}e^{-\frac{x^2}{2\sigma^2}}, x>0\
\end{equation}
Given our communication signals are at high frequency, both shadow fade and multipath fade encompass high-frequency geographical data. Therefore, we have reason to believe that these noise elements can be isolated using a low-frequency filter.

\subsubsection{Properties of Signal Power Attenuation}
According to Sec.~\ref{sec:signal_loss} and Sec.~\ref{sec:local}, we have analyzed the primary causes and characteristics of signal degradation.

Concerning path propagation for signals above 3GHz, weather conditions are the primary factors for signal degradation. Rain and clouds, being large-scale, long-period environmental factors, exhibit some degree of signal strength continuity over short time intervals and in confined areas.

For terrestrial environments, three types of degradation can occur due to location: path loss, shadow fading, and multipath fading. Apart from the high-frequency signal of multipath fading, the mid-frequency signal from shadow fading (i.e., passed through a low pass filter) can also be assumed to exhibit certain continuity across space.

\emph{In general satellite service scenarios, a service that remains stable in both spatial and temporal domains (i.e., no significant changes in an short time or small range) is an important property for satellite networks in terms of consensus and objective service evaluation.}
This leads to an essential assumption for our next section, Sec.~\ref{sec:stp}.

\section{Decentralized Satellite Network Protocol}
\label{sec:stp}
In the context of satellite network architecture, various roles are played by numerous types of equipment, such as satellites, ground stations, commonly referred to as Satellite Ground Stations, receivers - exemplified by the 'Dish' in SpaceX's Starlink - and user devices including mobile phones and computers. The landscape of wireless communication incorporates scenarios as varied as the transmission of data from satellites to receivers, receivers functioning as Access Points (APs) to relay data to end devices, and bidirectional communication between ground stations and satellites. This paper introduces a comprehensive model for data transmission and reception, simplifying the different parties involved in communication into just two roles: sender and receiver. This fundamental model contributes to a broader applicability across various wireless communication scenarios.

\emph{In the context of an unreliable wireless communication service, we aim to construct a decentralized consensus protocol so that neither the transmitter nor the receiver is able to gain additional revenue through fraud in the network. }

We categorize the quantitative descriptions of communication services into two types:
\begin{enumerate}
\item The first type pertains to the state properties of the communication network service, such as whether the device is online, the bandwidth provided by the device, latency, signal coverage, and signal strength at the receiver, amongst others. Generally, observing these data over a period of $\mathcal{O}(T)$ is sufficient to obtain results.
\item The second type comprises accumulative data, with transmission service as a typical example. These data require cumulative computations over a certain period (i.e. Epoch) to be obtained.
\end{enumerate}
In this paper, we propose Proof of Distribution (PoD) to achieve consensus for the first type of data. Ensuring consensus in the first scenario, we employ Proof of Flow (PoF) to obtain consensus in the second scenario.

\subsection{Spacetime Consistent Continuity Hypothesis}
\label{sec:space_time_con_hypo}
From the previous discussion on communication service scenarios, the import of the "Spatially Consistent Continuity Hypothesis" is that, for a communication network service, whether a fixed or mobile receiver is used, the service state parameters should not vary drastically due to minor spatial changes. If the range of state parameters can be bounded locally, it suffices to a certain extent to objectively evaluate the service quality of $\mathcal{T}$. In other words, we hope:
\begin{equation}
	\forall \mathcal{R}_1,\mathcal{R}_2 \in \left\{\mathcal{R}_\star\right\}:
	\lVert\mathbb{P}_1- \mathbb{P}_2\rVert<r \Rightarrow
	\lVert\alpha_{\mathbb{P}_1}-\alpha_{\mathbb{P}_2}\rVert<\delta_\alpha
\end{equation}
In this context, $\delta_\alpha$ denotes the high-frequency amplitude in the spatial distribution, while $r$ represents the radius of the service provided in $\mathcal{S}$. Worth noting is that, under general conditions, $r$ varies in different locations and is strictly less than the radius of the service offered by $\mathcal{S}$.

Despite the reality that local variations may be rather large due to situations such as shadowing or multipath effects discussed in Sec.~\ref{sec:local}, our discussion is not targeting a handful of local special cases. Instead, with the increase in sampling, and in compliance with the law of large numbers, the deviation from the mean should follow a Gaussian distribution. Let $x$ denote the distance between $\mathcal{P}_1$ and $\mathcal{P}_2$, i.e., $x:\lVert\mathbb{P}_1-\mathbb{P}_2\rVert$, we can then formulate the spatial continuity hypothesis:
\begin{hypothesis}[Spatially Consistent Continuity Hypothesis]
	\label{hypo:space}
	In a system composed of a transmitter and a receiver, barring any internal malicious activities, its local service state properties exhibit uniform consistency in the spatial domain:
	\begin{equation}
		\label{equation:spatial_con}
		 \forall \mathcal{R}_1,\mathcal{R}_2 \in \left\{\mathcal{R}_\star\right\}:
		 x<r \Rightarrow
		 (\alpha_{\mathbb{P}_1}-\alpha_{\mathbb{P}_2}) \sim \mathcal{N}(x;0, \sigma_\alpha^2)
	 \end{equation}
	\end{hypothesis}

In terms of temporal stability, let's consider two distinct time points, $t$ and $t + \Delta t$. We aspire for the service state parameters not to undergo substantial changes over a brief period, which is to say that:
\begin{equation}
	\label{equ:temporal_ori_con}
	 \forall \mathcal{R} \in \left\{\mathcal{R}_\star\right\}:
	\lVert \alpha(t+\Delta t)-\alpha(t)\rVert <\delta^t_\alpha
 \end{equation}
 As $\alpha$ is expressed as a state parameter throughout the service. 
% we consider an instantaneous scenario: $\lim_{\Delta t\to 0}$. By arranging Eqn.~\ref{equ:temporal_ori_con}, 
 Similar to spatial hypothesis, we obtain the Temporally Consistent Continuity Hypothesis:

 \begin{hypothesis}[Temporally Consistent Continuity Hypothesis]
	\label{hypo:time}
	In a system composed of a transmitter and a receiver, in the absence of internal malicious activities, the state parameter of the receiver $\mathcal{R}$ exhibits uniform consistency in the temporal domain:
	\begin{equation}
		\label{equation:temporal_con}
		\forall \mathcal{R} \in \left\{\mathcal{R}_\star\right\}:
		\lVert\alpha(t)-\alpha(t+\Delta t)\rVert\sim \mathcal{N}(\Delta t;0, \sigma_t^2)
	 \end{equation}
	\end{hypothesis}
	that is, we consider this to be a Gaussian Process(GP).
\subsection{Proof of Distribution}
\label{sec:pod}
Consider for each transmitter $\mathcal{T}$, whether the evaluation data provided by the receiver $\mathcal{R}$ can be used to have an objective and quantifiable service evaluation for $\mathcal{T}$. Unfortunately, given that the current system is not reliable, it is not simply assumed that all data are reliable. We consider possible scenarios where an objective fraudulent cannot be made, for different situations. These scenarios could arise from the deliberate deception of $\mathcal{T}$ or $\mathcal{R}$, or from the decline in service quality due to the signal attenuation mentioned in Sec.~\ref{sec:signal_loss} and Sec.~\ref{sec:local}. Here we only consider the fraudulent of status parameters; for the fraudulent of accumulative parameters, we discuss it in Sec.~\ref{sec:pof}.

According to the space-time consistency hypothesis mentioned in Sec.\ref{sec:space_time_con_hypo}, all state parameters are consistently continuous in the local spatial domain and the local temporal domain. A preliminary idea is to ensure the consensus on service state parameters by confirming the parameter distribution of the service in the spatial or temporal domain through a consensus protocol in the local domain. That is to say, for a communication service of $\mathcal{T}$, by using Hypo.\ref{hypo:space} and Hypo.~\ref{hypo:time} as prior conditions, this ensures that all $\mathcal{R}$ receiving the service can reach consensus on the distribution of its service parameter $\alpha$.

Table ~\ref{table:fraudulent} shows the decision scenario of the tansmitter and the receiver. 

\begin{table}
	\label{table:fraudulent}
	\centering
	\caption{Scene of fraudulent situations between receivers and transmitters}
	\centering
	\begin{tabular}{cccccc}
		\toprule
		&&&\multicolumn{3}{c}{Transmitter $\mathcal{T}$}                   \\
		\cmidrule(r){4-6}
		&&&\multicolumn{2}{c}{Claim Trans.}     & Claim Not Trans.\\
		\cmidrule{4-5}
		&&&Actually Trans. & Actually Not Trans. & Actually Not Trans.\\
		\midrule
		Receiver $\mathcal{R}$ & Claim Recv.  & Actually Recv. &Consensus: Success&$\times$&$\times$\\
		\cmidrule{3-6}
		&&Actually Unrecv. &$\times$&Corporate Fraud&$\times$\\
		\cmidrule{2-6}
		&Claim Unrecv. & Actually Recv.&$\mathcal{R}$ Fraud&$\times$&$\times$\\
		\cmidrule{3-6}
		&& Actually Unrecv. &Objective Failure&$\mathcal{T}$ Fraud&Consensus: Failure\\
		\bottomrule
	\end{tabular}
	\label{tab:table}
\end{table}

For Proof of Distribution (PoD), considering the local consistency hypothesis, a consensus can be achieved if most receivers $\mathcal{R}$ submit state parameters of $\mathcal{T}$ faithfully. We discuss the following scenarios in detail:
\begin{itemize}
\item \textbf{$\mathcal{R}$ Fraud}: The transmitter $\mathcal{T}$ has actually transmitted data, but some malicious receivers, denoted as $\Set{\mathcal{R}_\mathcal{F}}$, claim they did not receive any data. Given the Spatially Consistent Hypothesis, the honest receivers near $\mathcal{R}_\mathcal{F}$, denoted as $\mathcal{R}_\mathcal{H}$, will submit the real state parameters. This way, we can ensure that the estimation of state parameters for honest $\mathcal{T}$ won't deviate significantly from the actual situation. Furthermore, if $\Set{\mathcal{R}_\mathcal{F}}$ consistently submits biased state hyperparameters, they would face penalties.
\item \textbf{$\mathcal{T}$ Fraud}: The transmitter $\mathcal{T}$ claims to have transmitted data, but it has not happened in reality. Then under the spatial consistency assumption, consensus with actual services occurring would not be achieved which effectively prevents $\mathcal{T}$ falsification of the actual service occurring.
\item \textbf{Objective Failure}: This scenario is similar to '$\mathcal{T}$ Fraud', where phenomena like rain attenuation could occur. There is no way to know whether the poor quality of service is due to fraud on the part of the transmitter $\mathcal{T}$ or due to objective reasons for signal attenuation. In this case, $\mathcal{T}$ cannot be penalized and the anomalous state parameter data must be discarded. On the other hand, the temporal homogeneity provides a low-pass filter in the time domain, effectively filtering out occasional service quality issues due to noise signals. To some extent, it is possible to distinguish between objective causes or subjective fraud scenarios.
\item \textbf{Corporate Fraud}: This scenario is essentially a combination of '$\mathcal{R}$ Fraud' and '$\mathcal{T}$ Fraud'. However, thanks to $\mathcal{R}_{\mathcal{H}}$ in the entire system, consensus on the service provided by $\mathcal{T}$ can be achieved. If a $\mathcal{T}_\mathcal{F}$ colludes with more than one $\mathcal{R}_\mathcal{F}$ to commit fraud, the number of $\mathcal{R}_\mathcal{H}$ willing to accept its services will diminish over time, rendering $\mathcal{T}_\mathcal{F}$ unprofitable.
\end{itemize}

If there is now a verifier $\mathcal{V^\star}$ in the service system who has access to the state parameters submitted by $\Set{\mathcal{R}_\star}$ against a particular $\mathcal{T}$. Then he can obtain the objective service situation of $\mathcal{T}$ through the Proof of Distribution(PoD) algorithm. But in reality, there will be some deviations due to fraud or system instability, and we need to evaluate and analyze these deviations. Proof of Distribution (PoD) defines a loss function to estimate the deviation of the whole system.

The total loss function consists of a loss in the time domain and a loss in the spatial domain.
\begin{equation}
	\mathcal{L}^\text{total} = \mathcal{L}^\text{space}+\mathcal{L}^\text{time}
	\end{equation}

Based on Eqn.~\ref{equation:spatial_con}, we can calculate the spatial loss function by obtaining the distance between the actual state parameter at the current position and the expected state parameter  estimated from the local spatial domain:
\begin{equation}
	\begin{aligned}
	\label{equ:space_loss}
	\mathcal{L}^\text{space} &= \gamma_s \int_{\Omega}\lVert\alpha(p)-\mathbb{E}\left[\alpha(p)\right]\rVert\text{d}p\\
	&=\gamma_s\int_{\Omega}\left\Vert\alpha(p)-\int_{\Omega}\alpha(q)\mathcal{N}(|p-q|;0,\sigma_\alpha^2)\text{d}q\right\Vert\text{d}p
	\end{aligned}
\end{equation}
It is worth noting that Eqn.~\ref{equ:space_loss} is of the form that defines the loss function and not the variance (although they are very similar)
Similarly, for the loss function in the time domain, there is:
\begin{equation}
\begin{aligned}
	\label{equ:time_loss}
	\mathcal{L}^\text{time} &= \gamma_t \int_{\Omega}\left\Vert\alpha(p,t)-\mathbb{E}\left[\alpha(p,t)\right]\right\Vert\text{d}p\\
	&=\gamma_t \int_{\Omega}\left\Vert\alpha(p,t)-\int_T \alpha(p,\tau)\mathcal{N}(|t-\tau|;0,\sigma_t^2)\text{d}\tau\right\Vert\text{d}p\\
	\end{aligned}
	\end{equation}
Combining Eqn.~\ref{equ:space_loss} and Eqn.~\ref{equ:time_loss}, we can get the total loss function for a particular location:

\begin{equation}
	\label{equ:pod_loss}
	\ell^\text{total} =\gamma_s\left\Vert\alpha(p)-\int_{\Omega}\alpha(q)\mathcal{N}(|p-q|;0,\sigma_\alpha^2)\text{d}q\right\Vert + \gamma_t \left\Vert\alpha(p,t)-\int_T \alpha(p,\tau)\mathcal{N}(|t-\tau|;0,\sigma_t^2)\text{d}\tau\right\Vert
	\end{equation}
Eqn.~\ref{equ:pod_loss} is the core loss function of PoD.
\subsection{Sampling and Discretization}

For Eqn.~\ref{equ:pod_loss}, we must consider the following points that may arise.
\begin{enumerate}
\item Due to the inability for the system to sample continuously in the time domain as well as the unidirectional nature of the time domain, the sampling period in general may be in the range of a few minutes or even an hour. So we simplify Eqn.~\ref{equ:time_loss} to compare only the gap between two frames so that the time loss function becomes:
\begin{equation}
	\ell_t=\gamma_t \lVert\alpha(t)-\alpha(t-\Delta t)\rVert
	\end{equation}
It is worth noting that $\ell_t$ is more like a regular term to ensure that the loss function is convex.
\item Spatial sampling produces a sampling bias due to the fact that $\mathcal{R}$ may be dense in physical space. 
% \todo{need survey}
The weights for each sampling point need to be calculated individually during the sampling process.
\item The absolute value of the $\gamma$ parameter is not important, but the ratio of $\gamma_s$ to $\gamma_t$ is more critical in determining the degree of penalization between time and space.
\item For parameter bias due to objective reasons and due to the presence of dishonest nodes in the system, the state parameters provided by these nodes may deviate significantly from the actual results. This deviation can seriously affect the evaluation of the system services, so we use two additional algorithms to solve this problem.
\begin{enumerate}
\item We use the EM algorithm\citeyearpar{EM1,EM} to define a discriminative model $\mathcal{D}$ that determines which values may not be 'honest' commits. In a blockchain scenario, however, too many iterations for algorithmic convergence can also affect the speed of zk-proofs and consensus, the solution is to generate $\mathcal{D}$ after a fixed number of iterations.
\item We use robust statistics\citeyearpar{huber2011robust} for data with relatively large deviations to minimize the impact of such data.
\end{enumerate}
\end{enumerate}
Optimizing and discretizing Eqn.~\ref{equ:pod_loss} with these points yields:

\begin{equation}
	\label{equ:pod_loss_dis}
	\ell^\text{total} =\gamma_s\left\Vert\rho_s\left(\alpha(p)-\sum_{q}\alpha(q)\mathcal{N}(|p-q|;0,\sigma_\alpha^2)\Psi(q)\right)\right\Vert + \gamma_t \lVert\rho_t\left(\alpha(t)-\alpha(t-\Delta t)\right)\rVert
	\end{equation}
where $\rho(\cdot)$ denotes maximum likelihood estimation (MLE) that uses robust statisics and EM algorithms, $\Psi$ is the weight of the sampling estimate..

\subsubsection{Robust Statistics}
\label{sec:robust}
For samples that deviate far from the distribution of the data, which can have a large impact on the estimation of the distribution, the main idea of robust statistics is to reduce the numerical weight of outliers that deviate from the distribution. In robust statistics this approach is called MLE like estimators. a classical M-estimator uses $\rho(x)=x$ if $x<|\theta_r|$ otherwise $0$, i.e. metric trimming, so that samples with deviations above $\theta_r$ are ignored. However, this method does not guarantee the continuity hypothesis at the boundary, so we use the soft trimming method: 
\begin{equation}
	\rho(x)=x\left[1-\left(\frac{x}{\theta_r}\right)^2\right]^2
\end{equation}
which ensures continuity at the boundary.

\subsubsection{Distributional Parameters Estimation}
\label{sec:em}
We use the expectation maximization (EM) algorithm for unsupervised learning to estimate the parameters of the $\alpha$ distribution. This approach is particularly important when dealing with models involving latent variables - latent variables that are not directly observed but inferred from other observed variables (here our hidden variables are the distribution parameters of the anomalous data). The algorithm runs through an alternating expectation (E) step, which computes the expectation of the latent variables from the current parameter estimates, and a maximization (M) step, which updates the parameter estimates based on the expectations computed in the E step. This iterative process not only helps to deal with incomplete datasets (where missing data are considered as latent data), but also helps to find local maxima of the likelihood (or posterior) function. The end result is a meaningful statistical model with the best parameter estimates based on the observed data.

In fraud scenarios, whose anomalous data distributions do not have a specific parameter and can only be approximated using non-parametric methods, we can use robust statistics to exclude these values. And for the possible rain attenuation, shadows, etc. in the path loss, we can estimate them using parametric methods with Gaussian or lognormal distributions. For simplicity we denote these parameters as $\theta_i$.

Another problem is that general unsupervised learning requires some prior conditions such as the number of clusters $n$, etc. Such a prior conditions are relatively easy to obtain in fraud-free scenarios, but fraudulent scenarios such data cannot be realistically obtained from $\mathcal{R}$ or $\mathcal{T}$. Our solution is that these prior conditions can be written on the chain by means of smart contracts and modified by DAO organizations. 

Since we have the time continuum hypothesis, each iteration can be initialized using $\theta_\star$ at $t-\Delta t$. The algorithm flow is as follows

\begin{itemize}
	\item $\mathbb{E}$ \textbf{step}: In this step, we use the current parameter $\theta_\star$ estimates to compute the expectation of the latent variables. That is, the data will be determined to be anomalous (whether fraud or rain failure, etc.) by the posterior probability.
   
	\item $\mathbb{M}$ \textbf{step}: maximize likelihood estimation, i.e., to estimate the value of $\theta_\star$ for the existing data classification for path loss, multipath, and fraud, and for the valid data, we calculate the mean of the temporal and spatial distributions in Eqn.~\ref{equ:pod_loss_dis} as:
	% \todo{need revision}
	\begin{equation}
		\begin{aligned}
   		\mu_s&=\sum_q\rho\left(\alpha(q),\mathcal{N}(|p-q|;0,\sigma^2),\theta_\star\right)\\
   \mu_t&=\rho\left(\alpha(t)-\alpha(t-\Delta t),\theta_\star\right)\\
		\end{aligned}
   \end{equation}
   and come as a valid $\alpha$. And for anomalous data, we estimate $\theta_\star$ such that
   \begin{equation}
   \hat{\theta}_\star = \argmax_{\theta_\star}\sum_i \log{p(x;\theta_i)}
   \end{equation}
   where $p$ can be Gaussian, Log-Gaussian, Rayleigh, etc., or it can be estimated using nonparametric methods.
\end{itemize}
These two steps are repeated until convergence, i.e., the difference between the parameter estimates from successive iterations becomes very small or the maximum number of iterations is reached.

\subsection{Workflow}
The flow of the verifier $\mathcal{V^\star}$ for the estimation of the state parameters of PoD is shown in Algorithm~\ref{alg:pod}.
\begin{algorithm}
	\caption{Proof of Distribution workflow}\label{alg:pod}
	\begin{algorithmic}[1]
		\State Intialization: $\mathcal{D}$, $\theta_\star$
		\For{each epoch}
			\State $\mathcal{V^\star}$ collects state parameters $\alpha$
			\State $\mathcal{V^\star}$ gets priori parameters $\mathcal{P}$ from the blockchain
			\Repeat 
				\State Discriminate the data with $\mathcal{D}(\theta_\star, \mathcal{P})$ and $\rho(\cdot)$
				\State Compute $\ell$ in Eqn.~\ref{equ:pod_loss_dis}
				\State Maximize likelihood estimation mentioned in Sec.~\ref{sec:em}
				\State Update $\mathcal{D}$, $\theta_\star$
			\Until {$\delta \ell<\theta_\epsilon$}
			\State Submit each $\mathcal{T}\mapsto\mathcal{R}$ state parameter $\alpha$ and loss $\ell$ to blockchain
		\EndFor
	\end{algorithmic}
  \end{algorithm}

\subsection{Proof of Mesh}
\label{sec:pom}
\begin{figure}[h] 
	\centering

	\tikzset{every picture/.style={line width=0.75pt}} %set default line width to 0.75pt        

	\begin{tikzpicture}[x=0.75pt,y=0.75pt,yscale=-1,xscale=1]
	%uncomment if require: \path (0,620); %set diagram left start at 0, and has height of 620
	
	%Rounded Rect [id:dp750124901638515] 
	\draw   (85.33,55.32) .. controls (85.33,51.41) and (88.51,48.24) .. (92.42,48.24) -- (140.25,48.24) .. controls (144.16,48.24) and (147.33,51.41) .. (147.33,55.32) -- (147.33,76.58) .. controls (147.33,80.49) and (144.16,83.67) .. (140.25,83.67) -- (92.42,83.67) .. controls (88.51,83.67) and (85.33,80.49) .. (85.33,76.58) -- cycle ;
	
	%Rounded Rect [id:dp48669235157220747] 
	\draw   (234,59.32) .. controls (234,55.41) and (237.17,52.24) .. (241.09,52.24) -- (288.91,52.24) .. controls (292.83,52.24) and (296,55.41) .. (296,59.32) -- (296,80.58) .. controls (296,84.49) and (292.83,87.67) .. (288.91,87.67) -- (241.09,87.67) .. controls (237.17,87.67) and (234,84.49) .. (234,80.58) -- cycle ;
	%Rounded Rect [id:dp18024763545264455] 
	\draw   (381.67,59.99) .. controls (381.67,56.08) and (384.84,52.9) .. (388.75,52.9) -- (436.58,52.9) .. controls (440.49,52.9) and (443.67,56.08) .. (443.67,59.99) -- (443.67,81.25) .. controls (443.67,85.16) and (440.49,88.33) .. (436.58,88.33) -- (388.75,88.33) .. controls (384.84,88.33) and (381.67,85.16) .. (381.67,81.25) -- cycle ;
	%Straight Lines [id:da22338939853916084] 
	\draw    (116,84) -- (116,536.68) ;
	%Straight Lines [id:da41093222066311164] 
	\draw    (265.33,88.67) -- (265.33,537.34) ;
	%Straight Lines [id:da6894658831569054] 
	\draw    (415.67,89.33) -- (415.33,470.68) ;
	%Rounded Rect [id:dp7656981854323481] 
	\draw   (532.67,240.99) .. controls (532.67,237.08) and (535.84,233.9) .. (539.75,233.9) -- (587.58,233.9) .. controls (591.49,233.9) and (594.67,237.08) .. (594.67,240.99) -- (594.67,262.25) .. controls (594.67,266.16) and (591.49,269.33) .. (587.58,269.33) -- (539.75,269.33) .. controls (535.84,269.33) and (532.67,266.16) .. (532.67,262.25) -- cycle ;
	%Straight Lines [id:da5685517166506151] 
	\draw    (563.33,270.1) -- (563.33,538.01) ;
	%Curve Lines [id:da4999285616148299] 
	\draw    (415.33,120.67) .. controls (383.65,83.05) and (382.03,166.31) .. (414.34,129.81) ;
	\draw [shift={(415.33,128.67)}, rotate = 130.28] [color={rgb, 255:red, 0; green, 0; blue, 0 }  ][line width=0.75]    (10.93,-3.29) .. controls (6.95,-1.4) and (3.31,-0.3) .. (0,0) .. controls (3.31,0.3) and (6.95,1.4) .. (10.93,3.29)   ;
	%Curve Lines [id:da7102646562981692] 
	\draw    (265,120.67) .. controls (233.32,83.05) and (231.7,166.31) .. (264.01,129.81) ;
	\draw [shift={(265,128.67)}, rotate = 130.28] [color={rgb, 255:red, 0; green, 0; blue, 0 }  ][line width=0.75]    (10.93,-3.29) .. controls (6.95,-1.4) and (3.31,-0.3) .. (0,0) .. controls (3.31,0.3) and (6.95,1.4) .. (10.93,3.29)   ;
	%Curve Lines [id:da43935587120006137] 
	\draw    (115.67,118.33) .. controls (83.99,80.71) and (82.36,163.97) .. (114.68,127.48) ;
	\draw [shift={(115.67,126.33)}, rotate = 130.28] [color={rgb, 255:red, 0; green, 0; blue, 0 }  ][line width=0.75]    (10.93,-3.29) .. controls (6.95,-1.4) and (3.31,-0.3) .. (0,0) .. controls (3.31,0.3) and (6.95,1.4) .. (10.93,3.29)   ;
	%Straight Lines [id:da5982542169828384] 
	\draw    (116.67,190.67) -- (264,153.33) ;
	%Straight Lines [id:da03420116014596131] 
	\draw    (267.33,196.67) -- (414.67,159.33) ;
	%Straight Lines [id:da18819121172344389] 
	\draw    (117.33,236) -- (264.67,198.67) ;
	%Straight Lines [id:da5661378591720216] 
	\draw    (117.33,270.67) -- (266,294.67) ;
	%Straight Lines [id:da09257263242216829] 
	\draw    (265.33,242.67) -- (414,266.67) ;
	%Curve Lines [id:da24509802940606518] 
	\draw    (115.67,248.33) .. controls (83.99,210.71) and (82.36,293.97) .. (114.68,257.48) ;
	\draw [shift={(115.67,256.33)}, rotate = 130.28] [color={rgb, 255:red, 0; green, 0; blue, 0 }  ][line width=0.75]    (10.93,-3.29) .. controls (6.95,-1.4) and (3.31,-0.3) .. (0,0) .. controls (3.31,0.3) and (6.95,1.4) .. (10.93,3.29)   ;
	%Curve Lines [id:da2699290274378463] 
	\draw    (265,312.33) .. controls (233.32,274.71) and (231.7,357.97) .. (264.01,321.48) ;
	\draw [shift={(265,320.33)}, rotate = 130.28] [color={rgb, 255:red, 0; green, 0; blue, 0 }  ][line width=0.75]    (10.93,-3.29) .. controls (6.95,-1.4) and (3.31,-0.3) .. (0,0) .. controls (3.31,0.3) and (6.95,1.4) .. (10.93,3.29)   ;
	%Curve Lines [id:da36631177916702273] 
	\draw    (414.33,354.33) .. controls (382.65,316.71) and (381.03,399.97) .. (413.34,363.48) ;
	\draw [shift={(414.33,362.33)}, rotate = 130.28] [color={rgb, 255:red, 0; green, 0; blue, 0 }  ][line width=0.75]    (10.93,-3.29) .. controls (6.95,-1.4) and (3.31,-0.3) .. (0,0) .. controls (3.31,0.3) and (6.95,1.4) .. (10.93,3.29)   ;
	%Curve Lines [id:da018475252484453808] 
	\draw  [dash pattern={on 4.5pt off 4.5pt}]  (266,482.68) .. controls (312.2,498.52) and (515.87,548.33) .. (562.63,521.51) ;
	\draw [shift={(564,520.68)}, rotate = 146.92] [color={rgb, 255:red, 0; green, 0; blue, 0 }  ][line width=0.75]    (10.93,-3.29) .. controls (6.95,-1.4) and (3.31,-0.3) .. (0,0) .. controls (3.31,0.3) and (6.95,1.4) .. (10.93,3.29)   ;
	%Straight Lines [id:da37555356040154786] 
	\draw    (414.67,436.01) -- (562,408.68) ;
	%Straight Lines [id:da6539780141619689] 
	\draw    (415.33,457.33) -- (482.67,468.01) -- (564,481.33) ;
	%Curve Lines [id:da39492506084598866] 
	\draw  [dash pattern={on 4.5pt off 4.5pt}]  (115.33,356.01) .. controls (145.85,416.37) and (335.42,461.56) .. (414.15,470.54) ;
	\draw [shift={(415.33,470.68)}, rotate = 186.34] [color={rgb, 255:red, 0; green, 0; blue, 0 }  ][line width=0.75]    (10.93,-3.29) .. controls (6.95,-1.4) and (3.31,-0.3) .. (0,0) .. controls (3.31,0.3) and (6.95,1.4) .. (10.93,3.29)   ;
	%Rounded Rect [id:dp8925287813687783] 
	\draw   (96.67,546.01) .. controls (96.67,541.59) and (100.25,538.01) .. (104.67,538.01) -- (578,538.01) .. controls (582.42,538.01) and (586,541.59) .. (586,546.01) -- (586,570.01) .. controls (586,574.43) and (582.42,578.01) .. (578,578.01) -- (104.67,578.01) .. controls (100.25,578.01) and (96.67,574.43) .. (96.67,570.01) -- cycle ;
	%Curve Lines [id:da28362865161757145] 
	\draw  [dash pattern={on 4.5pt off 4.5pt}]  (264.67,340.01) .. controls (310.87,355.85) and (521.72,291.98) .. (563.45,340.51) ;
	\draw [shift={(564.67,342.01)}, rotate = 233.01] [color={rgb, 255:red, 0; green, 0; blue, 0 }  ][line width=0.75]    (10.93,-3.29) .. controls (6.95,-1.4) and (3.31,-0.3) .. (0,0) .. controls (3.31,0.3) and (6.95,1.4) .. (10.93,3.29)   ;
	%Straight Lines [id:da31308191194164303] 
	\draw    (415.33,370.67) -- (564,394.67) ;
	
	% Text Node
	\draw (105.33,57.4) node [anchor=north west][inner sep=0.75pt]    {$G_{s}$};
	% Text Node
	\draw (254,61.4) node [anchor=north west][inner sep=0.75pt]    {$S_{\star }$};
	% Text Node
	\draw (408.33,62.07) node [anchor=north west][inner sep=0.75pt]    {$S$};
	% Text Node
	\draw (557.17,242.4) node [anchor=north west][inner sep=0.75pt]    {$R$};
	% Text Node
	\draw (314.5,114.23) node [anchor=north west][inner sep=0.75pt]  [font=\normalsize]  {$pk_{s} ,sk_{s} ,e_{s}$};
	% Text Node
	\draw (138.33,115.73) node [anchor=north west][inner sep=0.75pt]  [font=\normalsize]  {$pk_{s\star } ,sk_{s\star } ,e_{s\star }$};
	% Text Node
	\draw (9.5,114.07) node [anchor=north west][inner sep=0.75pt]  [font=\normalsize]  {$pk_{G} ,sk_{G} e_{G}$};
	% Text Node
	\draw (145.24,160.86) node [anchor=north west][inner sep=0.75pt]  [font=\normalsize,rotate=-345.35]  {$pk_{s\star } ,\ \{e_{s\star }\}_{pk}$};
	% Text Node
	\draw (300.07,165.86) node [anchor=north west][inner sep=0.75pt]  [font=\normalsize,rotate=-345.35]  {$pk_{s} ,\ \{e_{s}\}_{pk}$};
	% Text Node
	\draw (166.45,255.51) node [anchor=north west][inner sep=0.75pt]  [font=\normalsize,rotate=-9.67]  {$pk_{G} ,\ \{e_{G}\}_{pk}$};
	% Text Node
	\draw (305.14,224.64) node [anchor=north west][inner sep=0.75pt]  [font=\normalsize,rotate=-9.56]  {$pk_{s\star } ,\ \{e_{s\star }\}_{pk}$};
	% Text Node
	\draw (150.33,298.4) node [anchor=north west][inner sep=0.75pt]  [font=\normalsize]  {$H=f(\{e_{\star }\})$};
	% Text Node
	\draw (4.67,236.73) node [anchor=north west][inner sep=0.75pt]  [font=\normalsize]  {$H=f(\{e_{\star }\})$};
	% Text Node
	\draw (303,344.73) node [anchor=north west][inner sep=0.75pt]  [font=\normalsize]  {$H=f(\{e_{\star }\})$};
	% Text Node
	\draw (160.67,317.9) node [anchor=north west][inner sep=0.75pt]  [font=\normalsize]  {$k=\{H\}_{sk_{s\star }}^{-1}$};
	% Text Node
	\draw (10.83,258.07) node [anchor=north west][inner sep=0.75pt]  [font=\normalsize]  {$k=\{H\}_{sk_{G}}^{-1}$};
	% Text Node
	\draw (310.67,361.9) node [anchor=north west][inner sep=0.75pt]  [font=\normalsize]  {$k=\{H\}_{sk_{s}}^{-1}$};
	% Text Node
	\draw (218,550.34) node [anchor=north west][inner sep=0.75pt]  [font=\small] [align=left] {Broadcast $\displaystyle pk,\ \{e\}_{pk}$, and $\displaystyle S$ is online. };
	% Text Node
	\draw (478.21,362.13) node [anchor=north west][inner sep=0.75pt]  [font=\normalsize,rotate=-12.96] [align=left] {SYN};
	% Text Node
	\draw (456.85,410.43) node [anchor=north west][inner sep=0.75pt]  [font=\normalsize,rotate=-345.9] [align=left] {SYN-ACK};
	% Text Node
	\draw (435.9,439.59) node [anchor=north west][inner sep=0.75pt]  [font=\normalsize,rotate=-9.67] [align=left] {ACK(k, BlockHash)};
	% Text Node
	\draw (292.04,419.93) node [anchor=north west][inner sep=0.75pt]  [font=\normalsize,rotate=-13.06]  {$V_{k}( ACK)$};
	% Text Node
	\draw (397.36,488.01) node [anchor=north west][inner sep=0.75pt]  [font=\normalsize,rotate=-11.18]  {$V_{k}( ACK)$};
	% Text Node
	\draw (162.24,199.36) node [anchor=north west][inner sep=0.75pt]  [font=\normalsize,rotate=-345.35]  {$pk_{s} ,\ \{e_{s}\}_{pk}$};

	\end{tikzpicture}

\caption{Workflow of PoM}
\label{fig:PoM}
\end{figure}
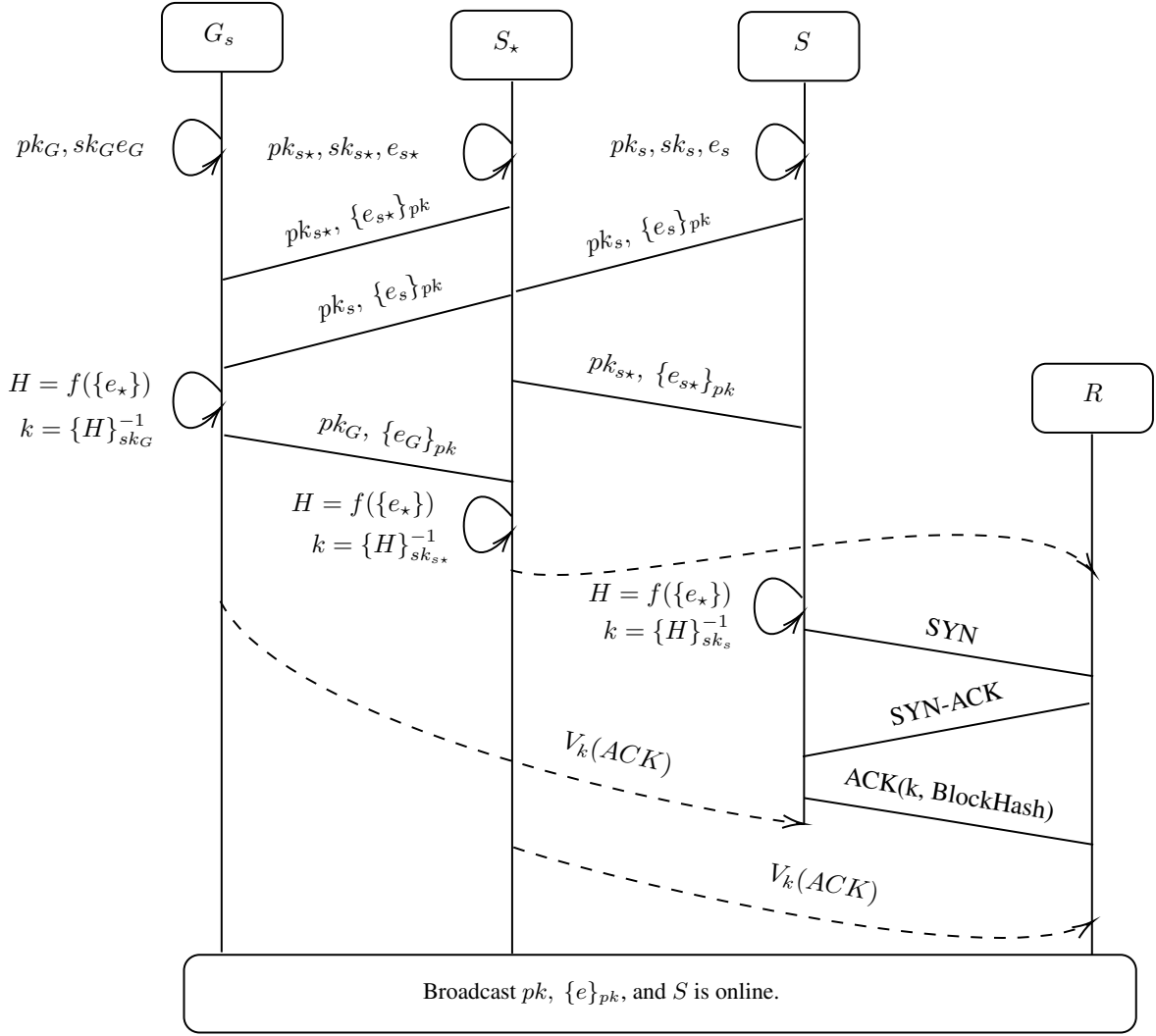

After a satellite successfully connects to the constellation, the first step it takes is to prove that it has indeed connected and begun delivering services. The node it carries initiates a transaction to register the satellite on the Space Network. Once ground stations $\mathcal{G}_S$, other satellites in the constellation $S_\star$ and $\mathcal{R}$ receive its online request, they begin the process of Meshing proofing. The satellite must provide a network connection proof to confirm certain service parameters before they can be written into the blockchain. Network connection proof involves communication among satellites, as well as between satellites and terminals and satellites and ground stations.
The process of proof of grid connection follows these stages:
\begin{enumerate}
	\item The satellite $\mathcal{S}$ generates a shared key $\mathcal{K}$ using Multi-key FHE (Fully Homomorphic Encryption) \citeyearpar{multikey_fhe}. The purpose of this step is to facilitate communication between the satellites  joining the constellation and with the ground station. Additionally, the generated key can be used in subsequent zk's \texttt{Gen} process to generate a key seed.
	\item The satellite participates in the PoD (Proof of Distribution) consensus with the ground nodes. The data transition layer in satellite-to-earth communication primarily uses the TCP/IP protocol. Upon connecting to the network, the satellite sends a \texttt{SYN} packet to the ground node. The ground node reciprocates by returning a \texttt{SYN-ACK} packet. The satellite then merges the current block hash with the \texttt{ACK} packet, signs it using $\mathcal{k}$, and sends it back to the ground node for verification.
	\item The ground station $\mathcal{G}_{S}$, constellation nodes $\mathcal{S}\star$, and recipient nodes $\mathcal{R}$ announce successful network connection to the satellite $\mathcal{S}$ on the blockchain.
\end{enumerate}
The whole process of network connection is illustrated in Fig.\ref{fig:PoM}.
\subsubsection{Multikey Fully Homomorphic Encryption}
We perform PoM using Multi-key FHE (fully homomorphic encryption).
\begin{enumerate}
	\item	Upon successfully connecting to the network physically, the satellite generates a private key $sk_{\mathcal{S}}$, a public key $pk_{\mathcal{S}}$, and redundant noise $e_{\mathcal{S}}$. Both $e$ and $sk$ are kept confidential.
	\item	The satellite uses $pk$ to encrypt $e$, yielding $\langle e\rangle_{pk}$.
	\item	The satellite broadcasts $pk$ and $\langle e\rangle_{pk}$ to $\mathcal{G}_\mathcal{S}$ and $\mathcal{S}_\star$
	\item $\mathcal{S}$ receives $pk_{\mathcal{G}}$ and $\langle e\rangle_{pk_\mathcal{S\star}}$ from the ground station and from the satellites in the same constellation.
	\item For all encrypted $e$, the satellite computes $\mathcal{H}({e})$, typically using a hash function.
	\item The satellite uses $sk_{\mathcal{S}}$ to decrypt $\mathcal{H}$ to get the final shared key $\mathcal{K}$.
\end{enumerate}
The process described above is similar to the Diffie-Hellman protocol but uses homomorphic encryption to generate a shared key based on $e_\star$. This ensures the entire protocol process is secure under at least the semi-honest model.
The PoM protocol is decribed in Algorithm~\ref{alg:pom}.
\begin{algorithm}
	\caption{Proof of Mesh workflow}\label{alg:pom}
	\begin{algorithmic}[1]
		\State Generate $sk_{\mathcal{S}}$, $pk_{\mathcal{S}}$, $e_{\mathcal{S}}$
		\State $\langle e_{\mathcal{S}}\rangle_{pk_s} \gets$\Call{Encrypt}{$e_{\mathcal{S}}$, $pk_{\mathcal{S}}$}
		\State Broadcast $pk_{\mathcal{S}}$, $\langle e_{\mathcal{S}}\rangle_{pk_s} $ to $\mathcal{G}_{\mathcal{S}}$, $\mathcal{S}_\star$
		\State Receive $pk_{\mathcal{S\star}}$, $\langle e_{\mathcal{S\star}}\rangle_{pk_{s\star}}$ from $\mathcal{S}_\star$
		\State Receive $pk_{\mathcal{G}}$, $\langle e_{\mathcal{G}}\rangle_{pk_G}$ from $\mathcal{G}_{\mathcal{S}}$
		\State $\mathcal{H} \gets \mathcal{G}(\langle e_{\mathcal{S\star}}\rangle_{pk_{s\star}},\langle e_{\mathcal{S}}\rangle_{pk_{s}},\langle e_{\mathcal{G}}\rangle_{pk_{G}})$
		\State $\mathcal{K} \gets \Call{Decrypt}{\mathcal{H}, sk_{\mathcal{S}}}$
		\State Send \texttt{SYN} to $\mathcal{R}$
		\State Waiting for \texttt{SYN-ACK} from $\mathcal{R}$
		\State \texttt{ACK} $\gets$ $\langle$\texttt{ACK} $||$ $\texttt{Hash}_\texttt{Block}$$\rangle_\mathcal{K}$
		\State $\mathcal{S}$ sends \texttt{ACK} to $\mathcal{R}$
	\end{algorithmic}
  \end{algorithm}
\subsection{Proof of Flow}
\label{sec:pof}
For cumulative data, such as the traffic used by a certain $\mathcal{R}$, it is not possible to use Proof of Distribution to prove that it is real data, since this type of data does not satisfy the spatio-temporal consistency assumption. 

A logical solution would be to record the signature in the header for every packet $\mathcal{T}$ and $\mathcal{R}$ transmitted, thus proving that $\mathcal{T}$ did indeed provide such a data message. There would be two problems with this scheme:
\begin{enumerate}
	\item There is no incentive for $\mathcal{R}$ to help $\mathcal{T}$ with signature verification. This is because it would require local computing resources for $\mathcal{R}$, resulting in $\mathcal{T}$ being inclined not to submit a signature but to claim that no actual service was received.
	\item  If the signature message is added to each message header, both for $\mathcal{R}$ and $\mathcal{T}$ invariably increase the computing pressure and network pressure.
\end{enumerate}

\begin{figure}
	\centering
	\includegraphics[width=1\textwidth]{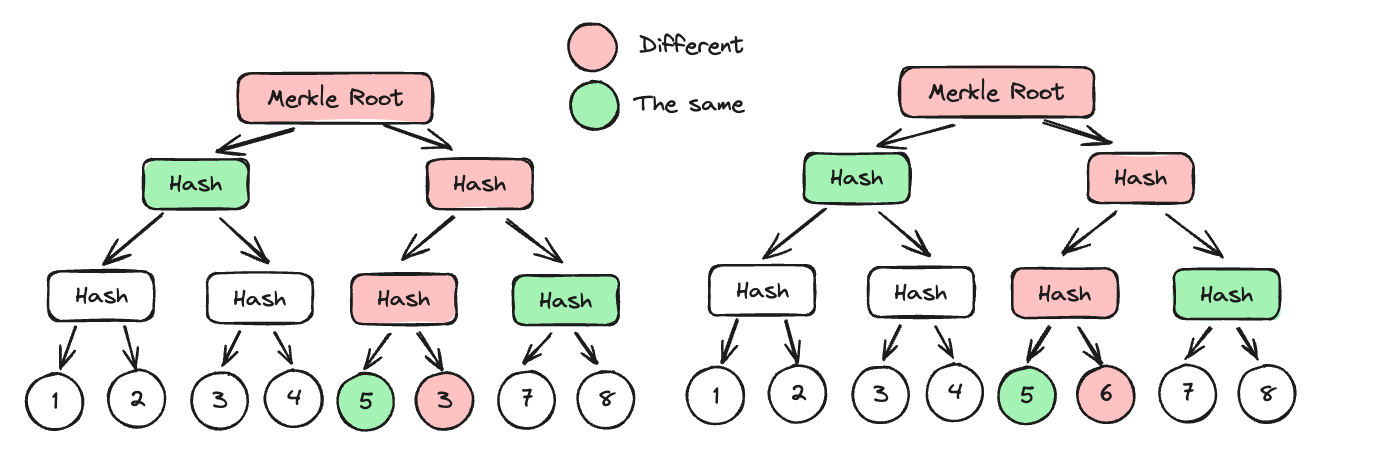}
	\caption{Compare two Merkle trees. In the best case it is $\mathcal{O}(1)$, in the worst case it is $\mathcal{O}(N)$, and if the number of errors in the messages is fixed, then the average case is $\mathcal{O}(\text{log}N)$. 
	So $\mathcal{R}$ cannot omit packets that were not transmitted successfully, but take their place as $\varnothing$ packets.}
	\label{fig:merkle}
\end{figure}

We propose several approaches to solve these two problems.
\begin{itemize}
	\item Instead of signing all messages, both $\mathcal{R}$ and $\mathcal{T}$ construct a Merkle tree\citeyearpar{becker2008merkle} for messages within a period of time $T$. This way only two Merkle trees need to be compared for comparison. Moreover, due to the nature of Merkle tree, it is possible to verify the network condition in parallel. As shown in Fig.~\ref{fig:merkle}
	\item Delayed commit: Delayed commit means that the Merkle tree will be submitted for messages from some point in the past, rather than real-time data. The significance of delayed submission is that since $\mathcal{R}$ and $\mathcal{T}$ have no way of knowing whether the current network service data may need to be submitted for proofs or not and must try to make traffic submissions without being nefarious.
	\item Two-Handshake: The process of two handshakes is that first $\mathcal{T}$ signs the Merkle tree structure and delivers it to $\mathcal{R}$, $\mathcal{R}$ signs it and delivers it to $\mathcal{T}$, and then $\mathcal{T}$ signs and submits it once more to $\mathcal{R} $ and the verifier $\mathcal{V^\star}$. The significance of the two handshakes is that if $\mathcal{T}$ discovers that $\mathcal{R}$ has not submitted a signature for a Merkle tree of sufficient size, then $\mathcal{R}$ can reject subsequent signatures and stop providing the service. Although in the short term this may lead to a reduction in revenue due to the rejection of signatures but in the long term it prevents $\mathcal{R}$ from cheating.

	\item $\mathcal{R}$ does not need to provide proofs continuously, but can submit proofs periodically in an Epoch. During the proof period, both $\mathcal{T}$ and $\mathcal{R}$ can submit Merkle trees for messages that have been signed 3 times (twice for $\mathcal{T}$ and once for $\mathcal{R}$).
	\item Thanks to PoD, we can know the approximate distribution of $\alpha$ for $\mathcal{T}$, which then requires $\mathcal{R}$ to submit a proof of whether the share of erroneous packets in the Merkle tree is consistent with the $\alpha$ of signaling stability obtained by PoD.
	Details are discussed in Sec.~\ref{sec:consensus}.
\end{itemize}

\subsection{Zero-Knowledge Proof}
Both PoD and PoF require the use of zkp to assist in submitting proofs to the blockchain. For PoD, the verifier $\mathcal{V}^\star$ needs to merge multiple $\mathcal{R}$ submissions of state parameters into a single proof, thus proving that $\mathcal{V}^\star$ did indeed compute the loss $\ell$. There are some cases where showing $\theta_\star$ is dangerous, because some $\mathcal{R}_\mathcal{F}$ may be able to reverse extrapolate the values they need to submit based on $\theta_\star$, so $\mathcal{V}^\star$ needs to show that it implements the Algorithm ~\ref{alg:pod} without shown without having to show $\theta^\star$.

We use zk-SNARKs\citeyearpar{zksnarks, groth2016size} to help verify PoD and PoF. zk-SNARKs is composed of a prover $\mathcal{P}$ and a verifier $\mathcal{V}$ (note the difference between $\mathcal{V}$ and the previously mentioned $\mathcal{V}^\star$), where the prover needs to provide a proof $\pi$ that he provided data services throughout the transmission. The prover can verify that the proof is valid within polynomial complexity via $\pi$ without needing to know $w$.
Mathematically, for a statement $x$ and a language $\mathcal{L}$, there exists a polynomial-complexity proof algorithm $A_\mathcal{L}$ that satisfies the following conditions:
\begin{equation}
\forall x \in \mathcal{L}, A_\mathcal{L}\in P(|x|): \exists w \Rightarrow A_\mathcal{L}(x,w)=1
\end{equation}. We can say $\mathcal{L}\in \text{NP}$.
The workflow of zk-SNARKs is described by a ternary operation:
[(\textrm{Gen}, \textrm{Prove}, \textrm{Verify})]
\begin{itemize}
\item \verb+Gen+: This is the initialization phase, which involves creating a "public parameter" $\text{crs}$ (public reference string), which will be used in the proof generation and verification phases, and which is known to both the prover $\mathcal{P}$ and the verifier $\mathcal{V^\star}$. Typically, this involves generating a pair of keys: a proof key and a verification key. Note that anyone can construct a proof using public parameters, but the specific information (i.e., the 'witness') remains hidden.
\item \verb+Prove+: in this step, the prover creates the proof $\pi$ using the proof key $pk$ from the setup phase and the instance-specific private input $w$.
\item \verb+Verify+: in this final step, the verifier uses the verification key $vk$ from the setup phase and the proof from the proof generation step to confirm that the prover does indeed have the knowledge they claim to have.
\end{itemize}

PLONK\citeyearpar{gabizon2019plonk} is the acronym for Permutations over Lagrange-bases for Oecumenical Noninteractive arguments of Knowledge. PLONK is a specific type of ZKP system designed for privacy and scalability. It is an ideal Zero Knowledge Proof protocol nowadays for the PoD network:

Here are the key points about PLONK:
\begin{itemize}
\item Non-Interactive: PLONK is a non-interactive ZKP system. This means that it requires only one message from the prover to the verifier, simplifying the communication process.
\item Succinctness: PLONK achieves remarkable succinctness in its proofs. The proofs are extremely short, making them computationally efficient to verify.
\item No Trusted Setup Ceremony: Unlike some other ZKP systems (such as zk-SNARKs), PLONK doesn’t require a specific trusted setup ceremony. This enhances security and privacy.
\item Universal Powers of Tau Ceremony: Instead of a dedicated phase 2 ceremony, PLONK relies on existing Hermez phase 1 ceremony files or the Perpetual Powers of Tau phase 1 output. These files serve as a universal foundation for PLONK circuits.
\item Gas Efficiency: The gas cost of a PLONK verifier is approximately 290k gas, compared to the 230k gas of Groth16 (another ZKP system).
\end{itemize}
PLONK strikes a balance between efficiency and security. So we use PLONK as our solution for zkp solution.

% \todo{something}
\subsubsection{Gaussian Distribution Approximation}
We use Halo2\citeyearpar{halo2} to implement the entire process of proof of PoD, since a Gaussian distribution is used in PoD, but it cannot be used directly in the zkp circuit, so it is necessary to approximate the Gaussian distribution by means of polynomial approximation.

We assume a Gaussian distribution $N(\mu;\sigma)$, the range of $\mathcal{R}$ is within $3\sigma$.
The kernel function is:
\begin{equation}
	\begin{aligned}
      W_{i,j} & = e^{-d^2_{i,j}/2\sigma^2}\\
\mathrm{Let}\ x &= d_{i,j}/\sigma\\
\text{then} \ W_{i,j} &= e^{-x^2/2}\\
	\end{aligned}
\end{equation}
A simple Taylor expansion of the Gaussian distribution kernel is:
\begin{equation}
	e^{-x^2/2} = \sum_{k=0}^\infty{\frac{2^{-k} (-x^2)^k}{k!}}
\end{equation}

Clearly, we cannot directly use an infinite Taylor expansion. Considering the range within $3\sigma$ and tolerance $\epsilon$, we propose:
\begin{equation}
	\begin{aligned}
	\frac{x^{2k}}{2^k k!} &< \epsilon \\
	2k\log{x}-k\log2-\log{k!} &< \log{\epsilon}
	\end{aligned}
\end{equation}
According to Stirling's formula, we have:
\begin{equation}
	\begin{aligned}
	\text{} k!&\sim \sqrt{2\pi k}\left(\frac{k}{e}\right)^k \\
	\log{k!}&=\frac{1}{2}[\log{2}+\log{\pi}+\log{k}+k*(\log{k}-1)]\\
	\end{aligned}
\end{equation}
According to our calculations, when $\epsilon = 10^{-5}$, it is probably sufficient to keep about 20 terms.

\subsubsection{MPC}
Since there is transaction prioritization on the blockchain, submissions for $\alpha$ or $\beta$ create situations where one can cheat by inferring from other nodes' $\alpha$ the $\alpha$ that best suits one's needs. Then we actually compute Eqn.~\ref{equ:pod_loss} while keeping $\alpha$ secret. 

There are many ways to implement Multi-Party Computing (MPC)\citeyearpar{damgaard2012multiparty,asharov2012multiparty,cramer2015secure}, which include garbled circuits\citeyearpar{bellare2012foundations} and TFHE (fully homomorphic encryption over the Torus)\citeyearpar{asharov2012multiparty}. We choose to use TFHE for multi-party computation.
\section{Consensus}
\label{sec:consensus}
This chapter focuses on the consensus framework of the Space Network. Sec.~\ref{sec:role} describes the roles involved in consensus. Next we describe how the roles collaborate with each other to accomplish the realization of the consensus network.

\begin{figure}[h] 
	\centering

	\tikzset{every picture/.style={line width=0.75pt}} %set default line width to 0.75pt        

	\begin{tikzpicture}[x=0.75pt,y=0.75pt,yscale=-1,xscale=1]
	%uncomment if require: \path (0,423); %set diagram left start at 0, and has height of 423
	
	%Flowchart: Alternative Process [id:dp5247219958287492] 
	\draw   (44.4,213.77) .. controls (44.4,210.69) and (46.89,208.2) .. (49.97,208.2) -- (108.83,208.2) .. controls (111.91,208.2) and (114.4,210.69) .. (114.4,213.77) -- (114.4,234.45) .. controls (114.4,237.52) and (111.91,240.01) .. (108.83,240.01) -- (49.97,240.01) .. controls (46.89,240.01) and (44.4,237.52) .. (44.4,234.45) -- cycle ;
	
	%Flowchart: Alternative Process [id:dp2733172774417989] 
	\draw   (181.4,213.77) .. controls (181.4,210.69) and (183.89,208.2) .. (186.97,208.2) -- (245.83,208.2) .. controls (248.91,208.2) and (251.4,210.69) .. (251.4,213.77) -- (251.4,234.45) .. controls (251.4,237.52) and (248.91,240.01) .. (245.83,240.01) -- (186.97,240.01) .. controls (183.89,240.01) and (181.4,237.52) .. (181.4,234.45) -- cycle ;
	%Straight Lines [id:da8837351661581803] 
	\draw    (114.4,222.8) -- (178.2,222.8) ;
	\draw [shift={(180.2,222.8)}, rotate = 180] [color={rgb, 255:red, 0; green, 0; blue, 0 }  ][line width=0.75]    (10.93,-3.29) .. controls (6.95,-1.4) and (3.31,-0.3) .. (0,0) .. controls (3.31,0.3) and (6.95,1.4) .. (10.93,3.29)   ;
	%Flowchart: Alternative Process [id:dp9868340888495897] 
	\draw  [color={rgb, 255:red, 0; green, 0; blue, 0 }  ,draw opacity=1 ] (318.4,212.77) .. controls (318.4,209.69) and (320.89,207.2) .. (323.97,207.2) -- (382.83,207.2) .. controls (385.91,207.2) and (388.4,209.69) .. (388.4,212.77) -- (388.4,233.45) .. controls (388.4,236.52) and (385.91,239.01) .. (382.83,239.01) -- (323.97,239.01) .. controls (320.89,239.01) and (318.4,236.52) .. (318.4,233.45) -- cycle ;
	
	%Straight Lines [id:da4012919981419476] 
	\draw [color={rgb, 255:red, 0; green, 0; blue, 255 }  ,draw opacity=1 ]   (253.33,223.8) -- (315.7,223.8) ;
	\draw [shift={(317.7,223.8)}, rotate = 180] [color={rgb, 255:red, 0; green, 0; blue, 255 }  ,draw opacity=1 ][line width=0.75]    (10.93,-3.29) .. controls (6.95,-1.4) and (3.31,-0.3) .. (0,0) .. controls (3.31,0.3) and (6.95,1.4) .. (10.93,3.29)   ;
	%Flowchart: Alternative Process [id:dp12429651334341663] 
	\draw  [color={rgb, 255:red, 0; green, 0; blue, 0 }  ,draw opacity=1 ] (484.9,212.27) .. controls (484.9,209.19) and (487.39,206.7) .. (490.47,206.7) -- (549.33,206.7) .. controls (552.41,206.7) and (554.9,209.19) .. (554.9,212.27) -- (554.9,232.95) .. controls (554.9,236.02) and (552.41,238.51) .. (549.33,238.51) -- (490.47,238.51) .. controls (487.39,238.51) and (484.9,236.02) .. (484.9,232.95) -- cycle ;
	%Curve Lines [id:da5586995417559117] 
	\draw  [dash pattern={on 4.5pt off 4.5pt}]  (99,239.84) .. controls (114.1,353.45) and (250.81,352.55) .. (287.08,350.43) ;
	\draw [shift={(288.67,350.33)}, rotate = 176.32] [color={rgb, 255:red, 0; green, 0; blue, 0 }  ][line width=0.75]    (10.93,-3.29) .. controls (6.95,-1.4) and (3.31,-0.3) .. (0,0) .. controls (3.31,0.3) and (6.95,1.4) .. (10.93,3.29)   ;
	%Curve Lines [id:da7823222977589079] 
	\draw [color={rgb, 255:red, 0; green, 0; blue, 255 }  ,draw opacity=1 ] [dash pattern={on 4.5pt off 4.5pt}]  (218.5,205.8) .. controls (232.98,144.38) and (277.63,119.26) .. (318.79,120.91) ;
	\draw [shift={(320.67,121)}, rotate = 183.43] [color={rgb, 255:red, 0; green, 0; blue, 255 }  ,draw opacity=1 ][line width=0.75]    (10.93,-3.29) .. controls (6.95,-1.4) and (3.31,-0.3) .. (0,0) .. controls (3.31,0.3) and (6.95,1.4) .. (10.93,3.29)   ;
	%Curve Lines [id:da5525214826214746] 
	\draw    (388.4,212.77) .. controls (405.95,198.39) and (466.75,195.89) .. (483.69,211.06) ;
	\draw [shift={(484.9,212.27)}, rotate = 227.73] [color={rgb, 255:red, 0; green, 0; blue, 0 }  ][line width=0.75]    (10.93,-3.29) .. controls (6.95,-1.4) and (3.31,-0.3) .. (0,0) .. controls (3.31,0.3) and (6.95,1.4) .. (10.93,3.29)   ;
	%Curve Lines [id:da05870903146514439] 
	\draw    (389.94,234.93) .. controls (410.85,253.66) and (465.99,251.77) .. (484.9,232.95) ;
	\draw [shift={(388.4,233.45)}, rotate = 46.01] [color={rgb, 255:red, 0; green, 0; blue, 0 }  ][line width=0.75]    (10.93,-3.29) .. controls (6.95,-1.4) and (3.31,-0.3) .. (0,0) .. controls (3.31,0.3) and (6.95,1.4) .. (10.93,3.29)   ;
	%Shape: Ellipse [id:dp8224802545412777] 
	\draw   (288.67,350.33) .. controls (288.67,339.29) and (304.34,330.33) .. (323.67,330.33) .. controls (343,330.33) and (358.67,339.29) .. (358.67,350.33) .. controls (358.67,361.38) and (343,370.33) .. (323.67,370.33) .. controls (304.34,370.33) and (288.67,361.38) .. (288.67,350.33) -- cycle ;
	
	%Curve Lines [id:da7053068472151311] 
	\draw  [dash pattern={on 4.5pt off 4.5pt}]  (358.67,350.33) .. controls (449.21,342.54) and (584.97,300.93) .. (549.87,239.44) ;
	\draw [shift={(549.33,238.51)}, rotate = 59.17] [color={rgb, 255:red, 0; green, 0; blue, 0 }  ][line width=0.75]    (10.93,-3.29) .. controls (6.95,-1.4) and (3.31,-0.3) .. (0,0) .. controls (3.31,0.3) and (6.95,1.4) .. (10.93,3.29)   ;
	%Shape: Ellipse [id:dp6488224812961816] 
	\draw  [color={rgb, 255:red, 0; green, 0; blue, 255 }  ,draw opacity=1 ] (320.67,121) .. controls (320.67,109.95) and (336.34,101) .. (355.67,101) .. controls (375,101) and (390.67,109.95) .. (390.67,121) .. controls (390.67,132.05) and (375,141) .. (355.67,141) .. controls (336.34,141) and (320.67,132.05) .. (320.67,121) -- cycle ;
	%Curve Lines [id:da1373106313170731] 
	\draw  [dash pattern={on 4.5pt off 4.5pt}]  (373.5,138.8) .. controls (388.13,162.2) and (375.18,183.7) .. (371.75,205.15) ;
	\draw [shift={(371.5,206.8)}, rotate = 277.77] [color={rgb, 255:red, 0; green, 0; blue, 0 }  ][line width=0.75]    (10.93,-3.29) .. controls (6.95,-1.4) and (3.31,-0.3) .. (0,0) .. controls (3.31,0.3) and (6.95,1.4) .. (10.93,3.29)   ;
	%Flowchart: Multidocument [id:dp14889268485556206] 
	\draw  [fill={rgb, 255:red, 255; green, 255; blue, 255 }  ,fill opacity=1 ] (407.32,266.95) -- (457,266.95) -- (457,294.38) .. controls (425.95,294.38) and (432.16,304.27) .. (407.32,297.87) -- cycle ; \draw  [fill={rgb, 255:red, 255; green, 255; blue, 255 }  ,fill opacity=1 ] (401.11,271.1) -- (450.79,271.1) -- (450.79,298.53) .. controls (419.74,298.53) and (425.95,308.43) .. (401.11,302.03) -- cycle ; \draw  [fill={rgb, 255:red, 255; green, 255; blue, 255 }  ,fill opacity=1 ] (394.9,275.26) -- (444.58,275.26) -- (444.58,302.69) .. controls (413.53,302.69) and (419.74,312.58) .. (394.9,306.18) -- cycle ;
	
	%Curve Lines [id:da8324134582534763] 
	\draw [color={rgb, 255:red, 0; green, 0; blue, 255 }  ,draw opacity=1 ]   (497.32,238.05) .. controls (503.31,243.28) and (505.74,248.5) .. (505.74,253.38) .. controls (505.74,261.05) and (499.6,268.17) .. (490.39,273.54) .. controls (483.48,277.56) and (474.91,280.61) .. (466.34,282.22) .. controls (464.05,282.65) and (461.76,282.98) .. (465.41,282.39)(495.35,240.31) .. controls (500.46,244.77) and (502.74,249.17) .. (502.74,253.38) .. controls (502.74,260.21) and (496.91,266.27) .. (488.88,270.94) .. controls (482.24,274.81) and (474.01,277.73) .. (465.78,279.27) .. controls (463.58,279.69) and (461.38,280) .. (465.12,279.39) ;
	\draw [shift={(457.67,281.84)}, rotate = 355.91] [color={rgb, 255:red, 0; green, 0; blue, 255 }  ,draw opacity=1 ][line width=0.75]    (10.93,-3.29) .. controls (6.95,-1.4) and (3.31,-0.3) .. (0,0) .. controls (3.31,0.3) and (6.95,1.4) .. (10.93,3.29)   ;
	%Curve Lines [id:da4239290977335963] 
	\draw    (350.44,240.92) .. controls (349.63,243.82) and (349.25,246.5) .. (349.25,248.97) .. controls (349.25,258.73) and (355.12,265.23) .. (362.13,269.3) .. controls (364.05,270.42) and (366.05,271.35) .. (368.05,272.12) .. controls (368.96,272.48) and (369.88,272.79) .. (370.79,273.08) .. controls (381.22,276.37) and (390.1,278.68) .. (388.47,278.26)(347.56,240.1) .. controls (346.66,243.3) and (346.25,246.25) .. (346.25,248.97) .. controls (346.25,259.92) and (352.72,267.3) .. (360.62,271.9) .. controls (362.68,273.09) and (364.82,274.09) .. (366.97,274.92) .. controls (367.94,275.3) and (368.92,275.64) .. (369.88,275.94) .. controls (380.35,279.25) and (389.27,281.56) .. (387.64,281.14) ;
	\draw [shift={(395.67,281.84)}, rotate = 199.13] [color={rgb, 255:red, 0; green, 0; blue, 0 }  ][line width=0.75]    (10.93,-3.29) .. controls (6.95,-1.4) and (3.31,-0.3) .. (0,0) .. controls (3.31,0.3) and (6.95,1.4) .. (10.93,3.29)   ;
	%Curve Lines [id:da6447806451923059] 
	\draw  [dash pattern={on 4.5pt off 4.5pt}]  (511.69,204.6) .. controls (500.67,180.37) and (418.06,141.54) .. (387.67,131.84) ;
	\draw [shift={(512.5,206.8)}, rotate = 254.48] [color={rgb, 255:red, 0; green, 0; blue, 0 }  ][line width=0.75]    (10.93,-3.29) .. controls (6.95,-1.4) and (3.31,-0.3) .. (0,0) .. controls (3.31,0.3) and (6.95,1.4) .. (10.93,3.29)   ;
	%Curve Lines [id:da103649136600382] 
	\draw [color={rgb, 255:red, 0; green, 0; blue, 255 }  ,draw opacity=1 ] [dash pattern={on 4.5pt off 4.5pt}]  (342.54,142.72) .. controls (334.1,162.51) and (325.38,180.23) .. (335.99,204.67) ;
	\draw [shift={(336.67,206.18)}, rotate = 245.22] [color={rgb, 255:red, 0; green, 0; blue, 255 }  ,draw opacity=1 ][line width=0.75]    (10.93,-3.29) .. controls (6.95,-1.4) and (3.31,-0.3) .. (0,0) .. controls (3.31,0.3) and (6.95,1.4) .. (10.93,3.29)   ;
	\draw [shift={(343.33,140.84)}, rotate = 112.85] [color={rgb, 255:red, 0; green, 0; blue, 255 }  ,draw opacity=1 ][line width=0.75]    (10.93,-3.29) .. controls (6.95,-1.4) and (3.31,-0.3) .. (0,0) .. controls (3.31,0.3) and (6.95,1.4) .. (10.93,3.29)   ;
	%Flowchart: Alternative Process [id:dp5844821272538874] 
	\draw  [color={rgb, 255:red, 0; green, 0; blue, 0 }  ,draw opacity=1 ] (109.4,65.77) .. controls (109.4,62.69) and (111.89,60.2) .. (114.97,60.2) -- (173.83,60.2) .. controls (176.91,60.2) and (179.4,62.69) .. (179.4,65.77) -- (179.4,86.45) .. controls (179.4,89.52) and (176.91,92.01) .. (173.83,92.01) -- (114.97,92.01) .. controls (111.89,92.01) and (109.4,89.52) .. (109.4,86.45) -- cycle ;
	%Curve Lines [id:da8829442373990062] 
	\draw  [dash pattern={on 4.5pt off 4.5pt}]  (142.5,92.8) .. controls (102.9,128.44) and (75.06,149.38) .. (83.24,207.05) ;
	\draw [shift={(83.5,208.8)}, rotate = 261.33] [color={rgb, 255:red, 0; green, 0; blue, 0 }  ][line width=0.75]    (10.93,-3.29) .. controls (6.95,-1.4) and (3.31,-0.3) .. (0,0) .. controls (3.31,0.3) and (6.95,1.4) .. (10.93,3.29)   ;
	%Curve Lines [id:da4603776977919758] 
	\draw  [dash pattern={on 4.5pt off 4.5pt}]  (153.5,91.8) .. controls (189.14,125.46) and (202.24,150.3) .. (211.23,208.04) ;
	\draw [shift={(211.5,209.8)}, rotate = 261.33] [color={rgb, 255:red, 0; green, 0; blue, 0 }  ][line width=0.75]    (10.93,-3.29) .. controls (6.95,-1.4) and (3.31,-0.3) .. (0,0) .. controls (3.31,0.3) and (6.95,1.4) .. (10.93,3.29)   ;
	%Curve Lines [id:da1858789086429029] 
	\draw  [dash pattern={on 4.5pt off 4.5pt}]  (185.85,76.32) .. controls (470.36,45.2) and (544.98,149.95) .. (539.68,206.12) ;
	\draw [shift={(539.5,207.8)}, rotate = 277.13] [color={rgb, 255:red, 0; green, 0; blue, 0 }  ][line width=0.75]    (10.93,-3.29) .. controls (6.95,-1.4) and (3.31,-0.3) .. (0,0) .. controls (3.31,0.3) and (6.95,1.4) .. (10.93,3.29)   ;
	\draw [shift={(181.5,76.8)}, rotate = 353.53] [color={rgb, 255:red, 0; green, 0; blue, 0 }  ][line width=0.75]    (10.93,-3.29) .. controls (6.95,-1.4) and (3.31,-0.3) .. (0,0) .. controls (3.31,0.3) and (6.95,1.4) .. (10.93,3.29)   ;
	%Rounded Rect [id:dp2245016284026704] 
	\draw  [color={rgb, 255:red, 0; green, 0; blue, 0 }  ,draw opacity=1 ] (298,204.76) .. controls (298,197.71) and (303.71,192) .. (310.76,192) -- (566.74,192) .. controls (573.79,192) and (579.5,197.71) .. (579.5,204.76) -- (579.5,243.04) .. controls (579.5,250.09) and (573.79,255.8) .. (566.74,255.8) -- (310.76,255.8) .. controls (303.71,255.8) and (298,250.09) .. (298,243.04) -- cycle ;
	%Curve Lines [id:da8873984664761232] 
	\draw    (581.8,218.8) .. controls (621,189.4) and (610.64,107.19) .. (601.08,78.49) ;
	\draw [shift={(600.5,76.8)}, rotate = 70.24] [color={rgb, 255:red, 0; green, 0; blue, 0 }  ][line width=0.75]    (10.93,-3.29) .. controls (6.95,-1.4) and (3.31,-0.3) .. (0,0) .. controls (3.31,0.3) and (6.95,1.4) .. (10.93,3.29)   ;
	%Flowchart: Preparation [id:dp007466892255238111] 
	\draw   (552,56) -- (569.72,36) -- (628.78,36) -- (646.5,56) -- (628.78,76) -- (569.72,76) -- cycle ;
	%Curve Lines [id:da6148726701377023] 
	\draw    (552,56) .. controls (491.8,18) and (236.08,9.89) .. (174.74,59.45) ;
	\draw [shift={(173.83,60.2)}, rotate = 319.82] [color={rgb, 255:red, 0; green, 0; blue, 0 }  ][line width=0.75]    (10.93,-3.29) .. controls (6.95,-1.4) and (3.31,-0.3) .. (0,0) .. controls (3.31,0.3) and (6.95,1.4) .. (10.93,3.29)   ;
	%Flowchart: Alternative Process [id:dp2280309869434114] 
	\draw   (182.4,289.1) .. controls (182.4,286.03) and (184.89,283.53) .. (187.97,283.53) -- (246.83,283.53) .. controls (249.91,283.53) and (252.4,286.03) .. (252.4,289.1) -- (252.4,309.78) .. controls (252.4,312.85) and (249.91,315.35) .. (246.83,315.35) -- (187.97,315.35) .. controls (184.89,315.35) and (182.4,312.85) .. (182.4,309.78) -- cycle ;
	%Curve Lines [id:da388130150611006] 
	\draw  [dash pattern={on 4.5pt off 4.5pt}]  (108.83,240.01) .. controls (121.15,270.38) and (139.44,296.71) .. (179.49,301.32) ;
	\draw [shift={(181.33,301.51)}, rotate = 185.53] [color={rgb, 255:red, 0; green, 0; blue, 0 }  ][line width=0.75]    (10.93,-3.29) .. controls (6.95,-1.4) and (3.31,-0.3) .. (0,0) .. controls (3.31,0.3) and (6.95,1.4) .. (10.93,3.29)   ;
	%Curve Lines [id:da09533872251048714] 
	\draw  [dash pattern={on 4.5pt off 4.5pt}]  (254.67,300.84) .. controls (305.63,296.92) and (318.81,269.95) .. (331.87,242.52) ;
	\draw [shift={(332.67,240.84)}, rotate = 115.46] [color={rgb, 255:red, 0; green, 0; blue, 0 }  ][line width=0.75]    (10.93,-3.29) .. controls (6.95,-1.4) and (3.31,-0.3) .. (0,0) .. controls (3.31,0.3) and (6.95,1.4) .. (10.93,3.29)   ;
	
	% Text Node
	\draw (121.4,208) node [anchor=north west][inner sep=0.75pt]  [font=\scriptsize] [align=left] {Run Node};
	% Text Node
	\draw (61.8,214.6) node [anchor=north west][inner sep=0.75pt]   [align=left] {User};
	% Text Node
	\draw (267.9,207.5) node [anchor=north west][inner sep=0.75pt]  [font=\scriptsize] [align=left] {Stake};
	% Text Node
	\draw (324.8,213.6) node [anchor=north west][inner sep=0.75pt]   [align=left] {Validator};
	% Text Node
	\draw (425.73,205.67) node [anchor=north west][inner sep=0.75pt]  [font=\scriptsize] [align=left] {VDF};
	% Text Node
	\draw (496.3,213.6) node [anchor=north west][inner sep=0.75pt]   [align=left] {Leader};
	% Text Node
	\draw (152.83,336.83) node [anchor=north west][inner sep=0.75pt]   [align=left] {Tx.};
	% Text Node
	\draw (254.1,149.64) node [anchor=north west][inner sep=0.75pt]  [rotate=-328.01]  {$\alpha ,\beta $};
	% Text Node
	\draw (133.19,258.08) node [anchor=north west][inner sep=0.75pt]  [font=\scriptsize,rotate=-29.97] [align=left] {Delegate};
	% Text Node
	\draw (292.67,341) node [anchor=north west][inner sep=0.75pt]   [align=left] {Mempool};
	% Text Node
	\draw (490.17,327.5) node [anchor=north west][inner sep=0.75pt]   [align=left] {Tx.};
	% Text Node
	\draw (342,111) node [anchor=north west][inner sep=0.75pt]   [align=left] {$\displaystyle \mathcal{D}_{A}$};
	% Text Node
	\draw (384.98,158.44) node [anchor=north west][inner sep=0.75pt]  [rotate=-359.82]  {$\alpha ,\beta $};
	% Text Node
	\draw (413.33,281.33) node [anchor=north west][inner sep=0.75pt]   [align=left] {$\displaystyle \mathcal{B}$};
	% Text Node
	\draw (488.67,268.33) node [anchor=north west][inner sep=0.75pt]   [align=left] {{\scriptsize Proposal}};
	% Text Node
	\draw (342,275.67) node [anchor=north west][inner sep=0.75pt]   [align=left] {{\scriptsize Verify}};
	% Text Node
	\draw (459,143.07) node [anchor=north west][inner sep=0.75pt]    {$\pi $};
	% Text Node
	\draw (320,156) node [anchor=north west][inner sep=0.75pt]  [font=\normalsize] [align=left] {{\scriptsize $\displaystyle \pi $}};
	% Text Node
	\draw (422.4,230.33) node [anchor=north west][inner sep=0.75pt]  [font=\scriptsize] [align=left] {in Turn};
	% Text Node
	\draw (117.8,66.6) node [anchor=north west][inner sep=0.75pt]   [align=left] {Satellite};
	% Text Node
	\draw (61.65,148.12) node [anchor=north west][inner sep=0.75pt]  [rotate=-307.92] [align=left] {Service};
	% Text Node
	\draw (192.18,93.11) node [anchor=north west][inner sep=0.75pt]  [rotate=-59.61] [align=left] {Service};
	% Text Node
	\draw (421.44,59.24) node [anchor=north west][inner sep=0.75pt]  [rotate=-16.97] [align=left] {VDF Election};
	% Text Node
	\draw (560,48) node [anchor=north west][inner sep=0.75pt]   [align=left] {Space DAO};
	% Text Node
	\draw (341,33) node [anchor=north west][inner sep=0.75pt]   [align=left] {Control};
	% Text Node
	\draw (187.13,216.6) node [anchor=north west][inner sep=0.75pt]  [font=\footnotesize] [align=left] {Challenger};
	% Text Node
	\draw (259.62,279.83) node [anchor=north west][inner sep=0.75pt]  [font=\scriptsize,rotate=-332.89] [align=left] {Delegate};
	% Text Node
	\draw (186.8,292.6) node [anchor=north west][inner sep=0.75pt]  [font=\footnotesize] [align=left] {Delegator};
	% Text Node
	\draw (230.78,141.79) node [anchor=north west][inner sep=0.75pt]  [font=\scriptsize,rotate=-323.8] [align=left] {Stake/Slash};

	\end{tikzpicture}
	
\caption{Roles in the Space Network Consensus Framework.}
\label{fig:role}
\end{figure}
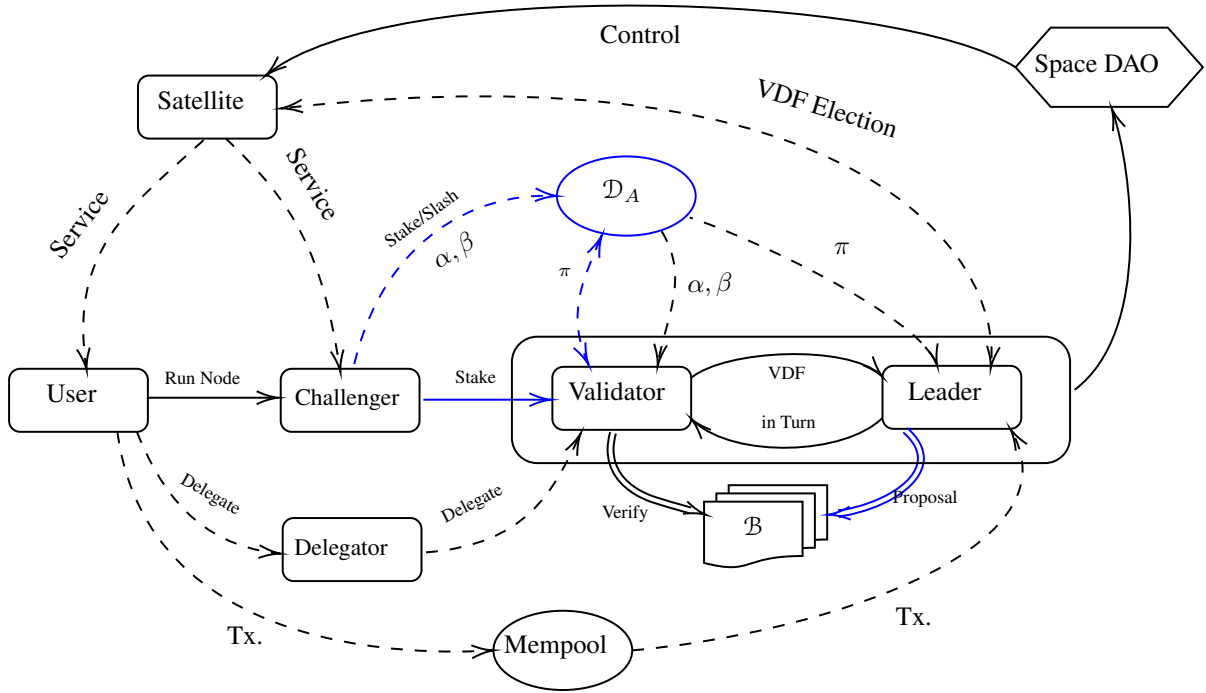

\subsection{Role}
\label{sec:role}

% \begin{figure}[h]
% 	\centering
% 	\includegraphics[width=0.8\textwidth]{images/miner_merkle.png}
% 	\caption{An example of a $\texttt{Tree}_{Merkle}$}
% 	\label{fig:miner_merkle}
% 	\end{figure}
There are several roles involved in consensus in Space Network.
\begin{itemize}
\item \texttt{User}:
The \texttt{User} receiving the network service, they can delegate \texttt{Validtor} to participate in mining and consensus, or they can become \texttt{Challenger} themselves to assist the network consensus.
\item \texttt{Challenger}:
Any \texttt{User} can become a \texttt{Challenger} by running the node program, they need to submit $\alpha$ and $\beta$ to the blockchain DA layer $\mathcal{D}_A$ in a time window $\varepsilon_m$.
\item \texttt{Validator}: The \texttt{Challenger} needs to stake \texttt{Token} to become a \texttt{Validator}, and they have two main responsibilities:
\begin{itemize}
\item Validates the voting consensus on \texttt{Leader}'s protosal Block $\mathcal{B}_p$.
\item Verify proofs of PoD and PoF submitted by \texttt{Leader}.
\end{itemize}
\item \texttt{Leader}
The system elects \texttt{Leaders} from among the \texttt{Validator} via VDF, and each \texttt{Leader} is responsible for the Proof-of-Distribution and Proof-of-Flow proofs of $\alpha$ and $\beta$ submitted by the \texttt{Challenger}, as well as the proposal of the block $\mathcal{B}$, in an Epoch $\varepsilon_i$.
\end{itemize}
The illustrative diagrams and flowcharts for the roles we show in Fig. ~\ref{fig:role}. We will explain each role next.

\subsection{Challenger}
\texttt{Challenger} needs to upload two types of data periodically,
%  \todo{or base on some Election?}
 one is the state parameter $\alpha$, which verifies the availability of the communication service, which is guaranteed by Proof-of-Distribution. Mines just need to submit $\alpha$ to $\mathcal{D}_A$ periodically to facilitate \texttt{Leader}'s use. If \texttt{Challenger} does not submit the state parameter $\mathcal{\alpha} $within this epoch $\mathcal{\varepsilon_m}$, then he will not be able to get the benefit of this epoch.

Another type of data that challengers need to submit is $\beta$, the traffic data of the communication service, which is guaranteed by Proof-of-Flow. The specific processing flow is as follows:
\begin{itemize}
\item $\mathcal{T}$ organizes the message into a Merkle tree $\texttt{Tree}_\texttt{Merkle}$ at $\varepsilon_m$ intervals and signs $\texttt{Tree}_\texttt{Merkle}$ as $\langle\texttt{Tree}_\texttt{Merkle}\rangle_{\mathrm{s}_\mathcal{T}}$ and send it to \texttt{Challenger}.
\item $\mathcal{R}$ verifies the Merkle tree (Figure~\ref{fig:merkle}) after receiving the message, and if the verification passes signs the root of the tree and resubmits it to $\mathcal{T}$. And store $\texttt{Tree}_\texttt{Merkle}$ in $\mathcal{D}_A$.
\item $\mathcal{T}$ signs the root of the tree and broadcasts it to all nodes $\texttt{Node}$ of the network.
In accordance with Sec.~\ref{sec:pof}, \texttt{Challenger} is required at $\varepsilon_m$. If either \texttt{Challenger} or $\mathcal{T}$ does not submit the traffic parameters within a limited time interval, then neither \texttt{Challenger} nor $\mathcal{R}$ will be able to gain this $\varepsilon_m$.
\end{itemize}
For specific workflow refer to Algorithm~\ref{alg:challenger} and Algorithm~\ref{alg:transmitter}

\begin{algorithm}[h]
	\label{alg:transmitter}
	\caption{Workflow of $\mathcal{T}$}
	\setstretch{1.3} 
\begin{algorithmic}[1]
	\Require $\mathcal{M}$, $\mathrm{s}_\mathcal{T}$
	\For{each $\varepsilon_m$}
	\State $\texttt{Tree}_\texttt{Merkle} \gets$ \Call{OrganizeMerkle}{$\mathcal{M}$}
	\State $\langle\texttt{Tree}_\texttt{Merkle}\rangle_{\mathrm{s}_\mathcal{T}} \gets$ \Call{Sign}{$\mathrm{s}_\mathcal{T}$, $\texttt{Tree}_\texttt{Merkle}$}
	\State \Call{Send}{$\mathcal{M}$, $\langle\texttt{Tree}_\texttt{Merkle}\rangle_{\mathrm{s}_\mathcal{T}}$, \texttt{Challenger}}
	\State \Call{Receive}{$\langle\texttt{Tree}_\texttt{Merkle}\rangle_{\mathrm{s}_\mathcal{T},\mathrm{s}_C}$, \texttt{Challenger}}
	\If{\Call{Verify}{$\mathrm{p}_C$, $\langle\texttt{Tree}_\texttt{Merkle}\rangle_{\mathrm{s}_\mathcal{T},\mathrm{s}_\mathcal{C}}$}}

	\State $\langle\texttt{Tree}_\texttt{Merkle}\rangle_{\mathrm{s}_\mathcal{T},\mathrm{s}_C, \mathrm{s}_\mathcal{T}} \gets $ \Call{Sign}{$\mathrm{s}_\mathcal{T}$, $\langle\texttt{Tree}_\texttt{Merkle}\rangle_{\mathrm{s}_\mathcal{T},\mathrm{s}_C}$}
	\State \Call{Broadcast}{$\langle\texttt{Tree}_\texttt{Merkle}\rangle_{\mathrm{s}_\mathcal{T},\mathrm{s}_C, \mathrm{s}_\mathcal{T}}$}
	\Else
		\State \Call{Drop}{$\langle\texttt{Tree}_\texttt{Merkle}\rangle_{\mathrm{s}_\mathcal{T},\mathrm{s}_C}$}
	\EndIf
	\EndFor
\end{algorithmic}
\end{algorithm}
	\begin{algorithm}[h]
		\label{alg:challenger}
		\setstretch{1.3} 
		\caption{Workflow of \texttt{Challenger}}
	\begin{algorithmic}[1]
		\For{each $\varepsilon_m$}
			\State \Call{Receive}{$\mathcal{M}$, $\langle\texttt{Tree}_\texttt{Merkle}\rangle_{\mathrm{s}_\mathcal{T}}$,$\mathcal{T}$}
			\State $\texttt{Tree}^C_\texttt{Merkle} \gets$ \Call{OrganizeMerkle}{$\mathcal{M}$} 
			\State $R_\mathcal{V} \gets $\Call{Verify}{$\mathrm{p}_\mathcal{T}$, $\langle\texttt{Tree}_\texttt{Merkle}\rangle_{\mathrm{s}_\mathcal{T}}$}
			
			\State $R_\mathcal{E} \gets$ \Call{Equal}{$\texttt{Tree}^C_\texttt{Merkle}$, $\texttt{Tree}_\texttt{Merkle}$}
			\If{$R_\mathcal{V}$ \textbf{and} $R_\mathcal{E}$}
			\State $\langle\texttt{Tree}_\texttt{Merkle}\rangle_{\mathrm{s}_\mathcal{T},\mathrm{s}_C} \gets$ \Call{Sign}{$\mathrm{s}_C$, $\langle\texttt{Tree}_\texttt{Merkle}\rangle_{\mathrm{s}_\mathcal{T}}$}
			\State \Call{Send}{$\langle\texttt{Tree}_\texttt{Merkle}\rangle_{\mathrm{s}_\mathcal{T},\mathrm{s}_C}$, $\mathcal{T}$}
			\State \Call{Receive}{$\langle\texttt{Tree}_\texttt{Merkle}\rangle_{\mathrm{s}_\mathcal{T},\mathrm{s}_C, \mathrm{s}_\mathcal{T}}$,$\mathcal{T}$}
			\State \Call{Store}{$\langle\texttt{Tree}_\texttt{Merkle}\rangle_{\mathrm{s}_\mathcal{T},\mathrm{s}_C, \mathrm{s}_\mathcal{T}}$, $\mathcal{D}_A$}
			\Else 
			\State  \Call{Drop}{$\mathcal{M}$}
			\EndIf
		\EndFor
	\end{algorithmic}
	\end{algorithm}

\subsection{Validator}
\texttt{User} can become \texttt{Validator} by taking tokens. Since \texttt{Validator} has the probability to become \texttt{Leader}, the hardware of \texttt{Validator} is a high configuration requirement. Because of this, we extend the delegated verification function at the contract layer. Users can delegate a certain \texttt{Validator} to assist them in validation (submitting $\alpha$, $\beta$ without having to stake \texttt{Token}.

\texttt{Validator} needs to validate $\alpha$ and $\beta$ that do sharding and submit $\pi$.

\texttt{Validtor} will be elected as \texttt{Leader} according to the blockchain system, and since this process is random, we discuss two blockchain random number generation schemes: the RANDAO and VDF

\subsubsection{RANDAO}
RANDAO (Random Number DAO) is a mechanism for producing decentralized, unbiased, unpredictable, and verifiable random numbers in a blockchain. It leverages the commitment scheme where each participant picks a number and keeps it secret, then publishes a hash of their secret number. The final random number is produced through a process that includes both commitments and reveals, minimizing the possibility for any participant to bias the outcome. Validators contribute randomness to the RANDAO with signatures over certain parameters, such as an epoch number 
\subsubsection{VDF}
Verifiable Delay Functions (VDFs) are a cryptographic primitive used in blockchains to ensure the security, transparency, and fairness of processes that require a certain amount of time to complete. This function takes significantly more time to compute than it does to verify, even on highly parallel processors. Its use cases include randomness generation, leader election, lotteries, and more in blockchain systems. The function works as an additional layer of security, deterring malicious activities that can influence certain outputs.

\subsubsection{Leader Election}
Each \texttt{Validator} and $\mathcal{T}$ can be elected as \texttt{Leader}, which we collectively refer to as \texttt{Validator} here. The PoS, PoD and PoF weights of \texttt{Validator} can be obtained from the contract. \texttt{Validator} determines the corresponding \texttt{Leader}  at each Epoch $\epsilon$ in accordance with the current state of the latest weight contract. Each \texttt{Leader} is responsible for packing $N_\varepsilon$ blocks in an $\epsilon$.

There will be a FIFO Leader Queue $\mathcal{Q}$ after system startup, and the system will first select $N_\mathcal{Q}$ counts of $\epsilon$ corresponding to \texttt{Leader} and store it in $\mathcal{Q}$. A sliding window is used to ensure that at least $N_\mathcal{Q}$ of \texttt{Leader} are in the current system. \texttt{Validator} needs to remove the current \texttt{Leader} from the queue after the current block is determined. \texttt{Validator} elects the new \texttt{Leader} at the same time and puts it into $\mathcal{Q}$. This approach effectively increases the blockchain throughput and isolates the authority to elect and pack blocks.

Each \texttt{Validator} uses the VDF algorithm as well as the current block hash $\mathcal{H}_{\mathcal{B}_i}$ to elect the \texttt{Leader} of the block $\mathcal{B}_{i+N_\mathcal{Q}}$ based on the various weighting data, and then push \texttt{Leader} into $\mathcal{Q}$.

\begin{figure}
	\centering
	\includegraphics[width=0.7\textwidth]{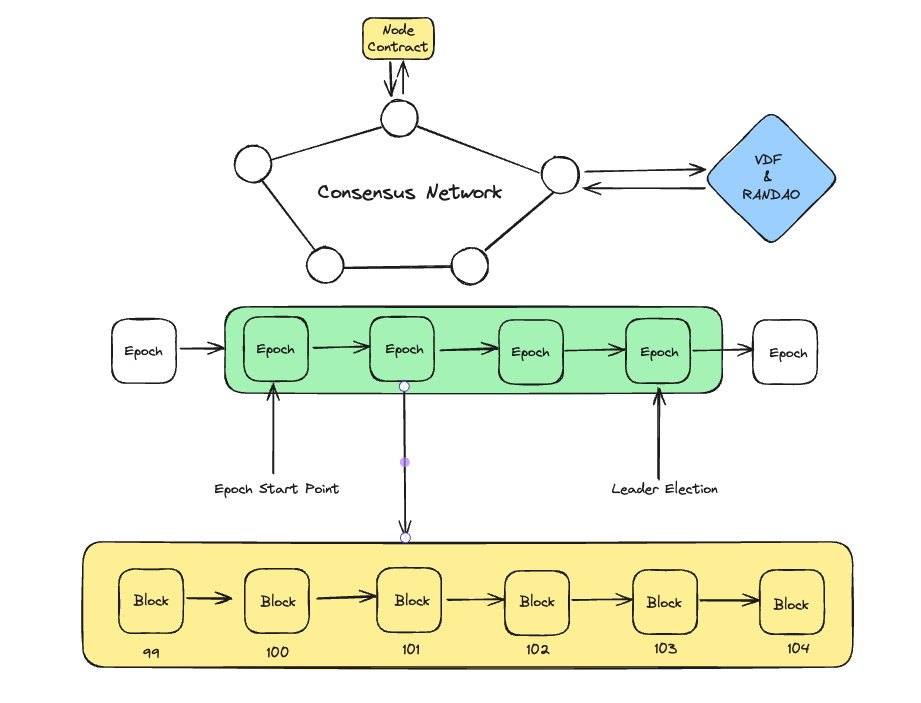}
	\caption{Description of the process by which \texttt{Validator} is elected as \texttt{Leader}.}
	\label{fig:election}
\end{figure}

Figure~\ref{fig:election} shows the Leader election process.

\subsection{Leader}
\texttt{Leader} is primarily responsible for the following duties:
\begin{itemize}
\item Submit proofs using algorithms ~\ref{alg:pod} and Sec.~\ref{sec:pod}.
\item Submit $\mathcal{B}_\mathcal{P}$
\item Put PoD and PoF proof as validation data to $\mathcal{B}$ and broadcast to \texttt{Validator} to validate and update weight contract.
\item If \texttt{Leader} has an exception such as downtime or network conditions, use ~\ref{sec:exception} to handle it.
\end{itemize}
Algorithm~\ref{alg:leader} illustrates the workflow of \texttt{Leader}

\begin{algorithm}[h]
	\label{alg:leader}
	\setstretch{1.3} 
	\caption{Workflow of \texttt{Leader}}
\begin{algorithmic}[1]
	\Require $N_\varepsilon$
	\For{each $\varepsilon$}
		\State $\alpha$,$\beta \gets$ \Call{Receive}{\texttt{Challenger}}
		\State $\pi_{\mathcal{D}} \gets$ \Call{PoD}{$\alpha$}
		\State $\pi_{\mathcal{F}} \gets$ \Call{PoF}{$\alpha$}
		\State \Call{Store}{$\pi_{\mathcal{D}}$,$\pi_{\mathcal{F}}$,$\mathcal{D}_A$}
		\State $\Delta S \gets$ \Call{Weight}{$\pi_{\mathcal{D}}$,$\pi_{\mathcal{F}}$}
		\For {$i \in {0..N_\varepsilon}$}
		\State $\mathcal{B}_\mathcal{P} = \varnothing$
		\While{\Call{Gas}{$\mathcal{B}_\mathcal{P}$}<$\mathcal{L}_{Gas}$ \textbf{and} $ \textproc{Mempool} \neq \varnothing $}
		\State $\uptau \gets$ \Call{SelectTx}{\textproc{Mempool}}
		\If {\Call{VerifyTx}{$\uptau$}}
		\State \Call{Put}{$\uptau$.,$\mathcal{B}_\mathcal{P}$}
		\EndIf
		\EndWhile
		\If{$i = N_\varepsilon - 1$}
		\State \Call{Put}{$\Delta S$, $\mathcal{B}_\mathcal{P}$}
		\EndIf
		\State $\mathcal{B}_\mathcal{P} \gets$ \Call{Sign}{$\mathcal{B}_\mathcal{P}$}
		\State \Call{Broadcast}{$\mathcal{B}_\mathcal{P}$}
		\EndFor
	\EndFor
\end{algorithmic}
\end{algorithm}

\begin{algorithm}[h]
	\label{alg:validator}
	\setstretch{1.3} 
	\caption{Workflow of \texttt{Validator}}
\begin{algorithmic}[1]
	\Require $\mathcal{Q}$,$N_\mathcal{Q}$
	\For{each $\varepsilon$}
		\While{$|\mathcal{Q}|<N_\mathcal{Q}$}
		\State \texttt{Leader} = \Call{Vdf}{ $\mathcal{W}$(\texttt{Validator}) }
		\State \Call{Put}{\texttt{Leader}, $\mathcal{Q}$}
		\EndWhile
		\State $\pi =$ \Call{Fetch}{$\mathcal{D}_A$}
		\If{\textbf{not} \Call{Verify}{$\pi$}}
		\State \Call{Exception}{\texttt{Leader}}
		\Else
			\For{$i \in 0..N_\varepsilon$}
				\State $\mathcal{B}_\mathcal{P} \gets$ \Call{ReceiveFrom}{\texttt{Leader}}
				\If{$\mathcal{B}_\mathcal{P} = \varnothing$}
				\State \Call{Exception}{\texttt{Leader}}
				\EndIf
				\For{$\uptau \in \mathcal{B}_\mathcal{P}$}
				\If{\textbf{not} \Call{VerifyTx}{$\uptau$}}
				\State \Call{Exception}{\texttt{Leader}}
				\EndIf
				\EndFor
				\If{\textbf{not} \Call{VerifyProof}{$\Delta S$, $\pi$}}
				\State \Call{Exception}{\texttt{Leader}}
				\EndIf
			\EndFor
		\EndIf
	\EndFor

\end{algorithmic}
\end{algorithm}

\begin{algorithm}[h]
	\label{alg:exception}
	\setstretch{1.3} 
	\caption{Exception Handling}
\begin{algorithmic}[1]
	\Procedure{Exception}{\texttt{Leader}}
	\If{\textbf{not} \Call{LeaderSkip}{\texttt{Leader}}}
	\State \Call{BroadcastLeaderSkip}{\texttt{Leader}}
	\EndIf
	\State \Call{StartTimer}{$\mathcal{T}_{R}$}
	\State \Call{VoteSkip}{\texttt{Leader}}
	\State \textsc{VoteCount} $\gets 1$
	\While{$\mathcal{T}_R \leq \mathcal{L}_R$ \textbf{and} \textsc{VoteCount} $ \leq 2/3 |\texttt{Validator}|$ }
	\State \textsc{VoteCount} $=$ \textsc{VoteCount} $+$ \Call{ReceiveSkipVote}{\text{Leader}}
	\EndWhile
	\If{$\mathcal{T}_R > \mathcal{L}_R$}
		\State \Return
	\Else
		\State \Call{Slash}{\texttt{Leader}}
		\State \Call{SkipLeader}{Leader}
	\EndIf
	\EndProcedure

\end{algorithmic}
\end{algorithm}

\subsubsection{Exception Handling}
\label{sec:exception}
\begin{figure}
	\centering
	\includegraphics[width=1\textwidth]{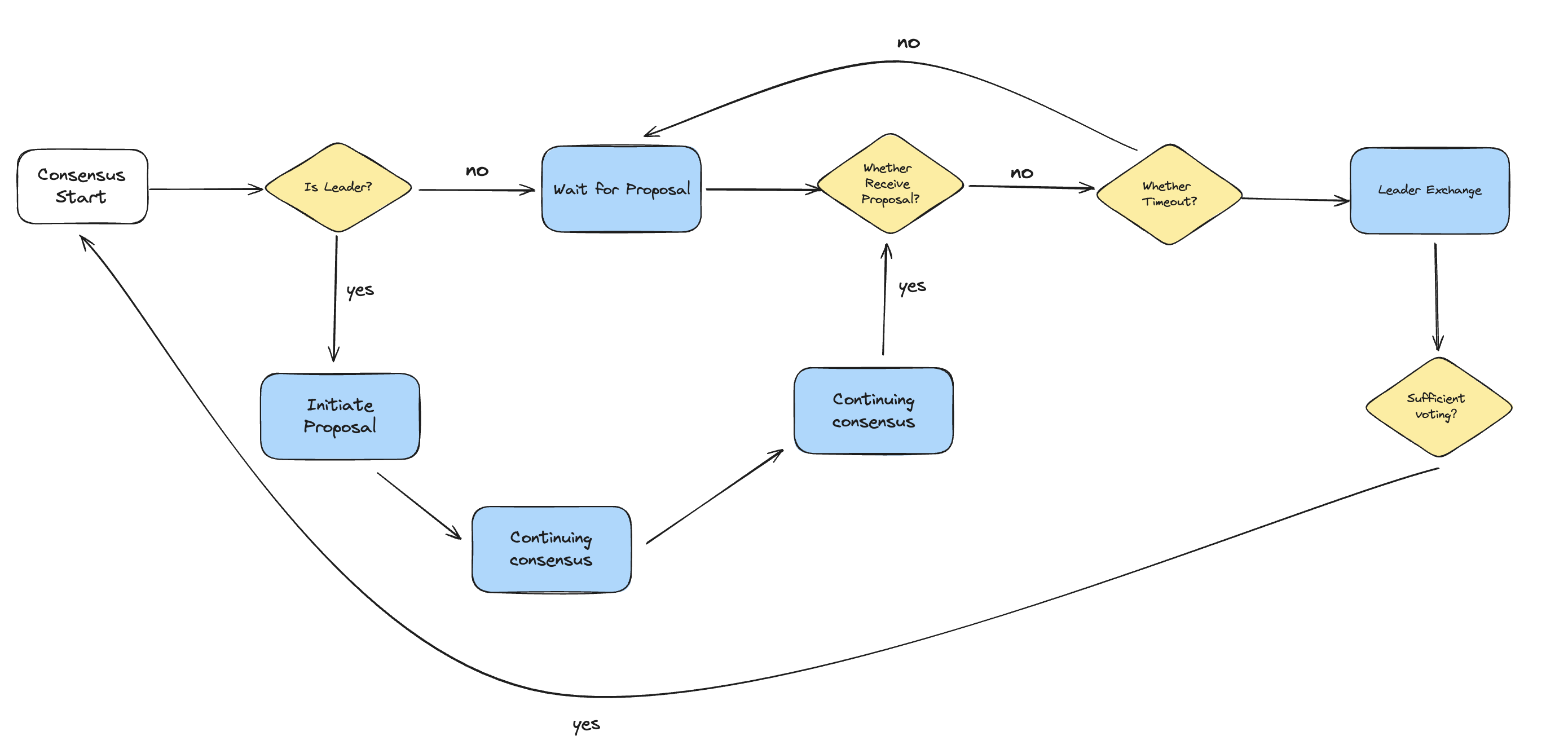}
	\caption{Exception handling.}
	\label{fig:exception}
\end{figure}

For scenarios where downtime or network problems prevent \texttt{Leader} from submitting blocks in time.
\begin{itemize}
\item \texttt{Validator} does not receive the leader proposal block at the proposal time, \texttt{Validator} initiates a vote to update \texttt{Leader} to other \texttt{Validator} to turn on the skip \texttt{Leader} timer.
\item All \texttt{Validator} check to see if $\mathcal{B}_\mathcal{P}$ has been received after receiving the vote to skip the \texttt{Leader}, and if it has not been received and the Proposal times out, record the request for a skip vote
\item \texttt{Validator} receives 2/3 of \texttt{Validator}'s votes before the skip \texttt{Leader} timer times out, completing the \texttt{Leader} skip, otherwise \texttt{Validator} looks at the height of the other blocks and if the block height continues to grow, then synchronization is turned on, otherwise the current height is considered to be an empty block.
\end{itemize}
Figure ~\ref{fig:exception} illustrates the exception handling process.

\subsection{Space DAO}
The Space DAO consists of \texttt{Validator} and \texttt{Leader}. Satellite parameters and orbits can be adjusted through contract control.

% \section{Application}
% \subsection{Contract}
% \begin{figure}
% 	\centering
% 	\includegraphics[width=0.7\textwidth]{images/contract1.png}
% 	\caption{Contract1}
% 	\label{fig:contract1}
% \end{figure}
% \begin{figure}
% 	\centering
% 	\includegraphics[width=0.7\textwidth]{images/contract2.png}
% 	\caption{Contract2}
% 	\label{fig:contract2}
% \end{figure}
% \begin{figure}
% 	\centering
% 	\includegraphics[width=0.7\textwidth]{images/contract3.png}
% 	\caption{Contract3}
% 	\label{fig:contract3}
% \end{figure}
\section{Result and Discussion}
\label{sec:result}
We conducted tests on multiple 4-core 16G AWS cloud platforms, using the Rust language for programming. Tests were carried out for Proof of Degrade (PoD), Proof of Failure (PoF), and Proof of Misbehavior (PoM).
\begin{itemize}
	\item \textbf{PoD}:  We conducted Proof of Degrade (PoD) tests for $N$ instances of $\mathcal{R}$ corresponding to one satellite. The result is shown in Table~\ref{tab:pod_result}.
	\item \textbf{PoF}: We conducted Proof of Failure (PoF) tests for $N$ instances of $\mathcal{R}$ corresponding to one satellite, with packet length of $L$. The result is shown in Table~\ref{tab:pof_result}.
	\item \textbf{PoM}: We used Zama's tfhe-rs\citeyearpar{TFHE-rs} for homomorphic encryption processing. The time taken for grid-connected private key exchange between two satellites is approximately 50 seconds.
	\item \textbf{Consensus}: We modified Tendermint from Cosmos to implement the whole process plan. Because of the asynchronous operation of Proof of Failure (PoF), Proof of Degrade (PoD), and Proof of Misbehavior (PoM), it only needs to produce results within 10 blocks to generate effective proofs and submit them to the blockchain.
\end{itemize}
\begin{table}[H]
	\centering
	\begin{minipage}{0.45\textwidth}
	  \centering
	  \caption{PoD Benchmark}
	  \label{tab:pod_result}
	  \begin{tabular}[t]{lc}
	  \hline
	  &Time\\
	  \hline
	  N=500&0.95s\\
	  N=1000&3.03s\\
	  N=2000&11.81s\\
	N=5000&76.27s\\
	  \hline
	  \end{tabular}
	\end{minipage}
	\hfill
	\begin{minipage}{0.45\textwidth}
	  \centering
	  \caption{PoF Benchmark}
	  \label{tab:pof_result}
	  \begin{tabular}[t]{lccc}
	  \hline
	  &L=128&L=256&L=512\\
	  \hline
	  N=500&0.29s&0.32s&0.36s\\
	  N=1000&0.56&0.61s&0.68s\\
	  N=5000&2.71s&3.03s&3.33s\\
	  N=10000&5.44s&5.87s&6.34s\\
	  \hline
	  \end{tabular}
	\end{minipage}
  \end{table}
\section{Conclusion}
We proposed a decentralized protocol based on satellite communication network services. It achieves consensus through decentralised proofs of satellite grid connections, satellite communication statuses and satellite-provided traffic via Proof of Misbehavior (PoM), Proof of Degrade (PoD), and Proof of Flow (PoF). It effectively prevents the transparency of communication network service evaluation.

% \section{Acknowledgment}
% {\color{red}{The method proposed in this paper is a lead-in approach. Unfortunately, due to the company's broken capital chain, the project cannot continue. If you are interested in this article or hoping for some opportunities for cooperation.", please feel free to send an email to the author at \href{mailto:shiotoli@gmail.com}{shiotoli@gmail.com}. Thank you for reading.}}
\bibliographystyle{unsrtnat}
\bibliography{references}  %%% Uncomment this line and comment out the ``thebibliography'' section below to use the external .bib file (using bibtex) .

%%% Uncomment this section and comment out the \bibliography{references} line above to use inline references.
% \begin{thebibliography}{1}

% 	\bibitem{kour2014real}
% 	George Kour and Raid Saabne.
% 	\newblock Real-time segmentation of on-line handwritten arabic script.
% 	\newblock In {\em Frontiers in Handwriting Recognition (ICFHR), 2014 14th
% 			International Conference on}, pages 417--422. IEEE, 2014.

% 	\bibitem{kour2014fast}
% 	George Kour and Raid Saabne.
% 	\newblock Fast classification of handwritten on-line arabic characters.
% 	\newblock In {\em Soft Computing and Pattern Recognition (SoCPaR), 2014 6th
% 			International Conference of}, pages 312--318. IEEE, 2014.

% 	\bibitem{keshet2016prediction}
% 	Keshet, Renato, Alina Maor, and George Kour.
% 	\newblock Prediction-Based, Prioritized Market-Share Insight Extraction.
% 	\newblock In {\em Advanced Data Mining and Applications (ADMA), 2016 12th International 
%                       Conference of}, pages 81--94,2016.

% \end{thebibliography}

% \appendix
% \section{Appendix A : Symbol Table}
% \label{app:A}

\end{document}